\documentstyle[epsf,amssymb,amsmath,ifthen,graphicx]{aa}
\def\astroncite#1#2{#2}
\def\aap{A\& A}
\def\araa{AnnRevA\& A}
\def\mnras{MNRAS}
\def\apj{ApJ}
\def\aj{AJ}
\def\apjl{ApJL}
\def\apjs{ApJS}
\def\jqsrt{JQSRT}

\def\bigegg{1}
\def\figs{1}

\def\incltables{1}

\def\figsctrans{0}
\def\NN{N}
\def\vec#1{{\mathbf{#1}}}

\def\sec{\hbox{s}}

\def\cm{\hbox{cm}}
\def\km{\hbox{km}}
\def\kel{\hbox{K}}

\def\yr{\hbox{yr}}
\def\AU{\hbox{AU}}

\def\saffine{s}
\def\bimpact{b}
\def\thinf{\Theta_{\infty}}

\def\testannulthin{2}
\def\solidang{{\boldsymbol\omega}}

\def\kin{\hbox{{\scriptsize kin}}}
\def\molec{\hbox{{\scriptsize mol}}}
\def\eff{\hbox{{\scriptsize eff}}}
\def\Hmolec{{\hbox{H}_2}}
\def\lwtot{a_{\hbox{\scriptsize tot}}}
\def\turblw{a_{\hbox{\scriptsize turb}}}
\def\innn{\hbox{{\scriptsize in}}}
\def\outtt{\hbox{{\scriptsize out}}}

\def\Raccr{{R_{\hbox{\scriptsize acc}}}}
\def\Rcentr{{R_{\hbox{\scriptsize c}}}}
\def\sech{\hbox{sech}}
\def\maxx{\hbox{\scriptsize max}}

\def\iter{\hbox{\scriptsize iter}}

\def\therm{\hbox{\scriptsize abs}}
\def\comma{\,,}
\def\fullstop{\,.}
\begin{document}
\thesaurus{12(02.18.7,02.12.3,08.03.4,08.06.2,13.09.6,13.20.1)}
\title{An efficient algorithm for two--dimensional radiative transfer in
axisymmetric circumstellar envelopes and disks} 
\author{C.P. Dullemond and R. Turolla} 
\authorrunning{Dullemond and Turolla}
\titlerunning{Two--dimensional radiative transfer in circumstellar envelopes
and disks} \institute{Max Planck Institut f\"ur Astrophysik, Karl
Schwarzschild Strasse 1,\\ D--85748 Garching, Germany e--mail:
dullemon@mpa-garching.mpg.de\\ Department of Physics, University of Padova,
Via Marzolo 8, 35131 Padova, Italy, e--mail: turolla@pd.infn.it}
\date{DRAFT, \today}

\maketitle

\begin{abstract}
We present an algorithm for two--dimensional radiative transfer in
axisymmetric, circumstellar media. The formal integration of the transfer
equation is performed by a generalization of the short characteristics (SC)
method to spherical coordinates. Accelerated Lambda Iteration (ALI) and Ng's
algorithm are used to converge towards a solution. By taking a
logarithmically spaced radial coordinate grid, the method has the natural
capability of treating problems that span several decades in radius, in the
most extreme case from the stellar radius up to parsec scale. Flux
conservation is guaranteed in spherical coordinates by a particular choice
of discrete photon directions and a special treatment of nearly--radially
outward propagating radiation. The algorithm works well from zero up to very
high optical depth, and can be used for a wide variety of transfer problems,
including non--LTE line formation, dust continuum transfer and high
temperature processes such as compton scattering. In this paper we focus on
multiple scattering off dust grains and on non-LTE transfer in molecular and
atomic lines. Line transfer is treated according to an ALI scheme for
multi-level atoms/molecules, and includes both random and systematic
velocity fields. The algorithms are implemented in a multi-purpose
user-friendly radiative transfer program named RADICAL. We present two
example computations: one of dust scattering in the Egg Nebula, and one of
non-LTE line formation in rotational transitions of HCO$^{+}$ in a flattened
protostellar collapsing cloud.
\end{abstract}

\begin{keywords}
radiative transfer -- line: profiles -- stars: formation, cirumstellar
matter -- submillimeter -- infrared: stars
\end{keywords}

\section{Introduction}
Molecular line and dust continuum observations are an important tool for
studying the envelopes and disks around young stellar objects (YSO),
post-AGB stars and AGN. One of the main difficulties in interpreting such
observations is that optical depths effects play an important role in the
emission of this radiation. It is, for instance, well known that
self-absorption and non-LTE effects are the main processes at work in
shaping the characteristic asymmetric double-peaked emission line profiles
from collapsing protostellar cores (Zhou, \cite{zhou:1992}). Radiative
transfer computations, at an appropriate level of complexity, are therefore
needed in order to reconstruct the density, velocity and temperature
structure of an observed cloud.

If densities drop below the critical density, the system may deviate from
local thermodynamic equilibrium (LTE). Line trapping and photon escape from
the line wings can in some cases be treated in the large-velocity-gradient
(LVG) limit, but this approach is only valid if systematic velocity fields
are much greater than the local turbulent line width. In all other cases
one must perform a full non-LTE line transfer computation. 

For problems that can be formulated in 1-D slab or spherical geometry there
exist many radiative transfer programs, many of which use sophisticated
techniques such as Accelerated Lambda Iteration (ALI; see a review by Hubeny
\cite{hubeny:1989}) and Complete Linearization (CL; Auer \& Mihalas
\cite{auermihalas:1969}). But the requirement of 1-D geometry is often too
restrictive. Distinctly non-spherical features are often observed from YSO,
such as bipolar reflection nebulae (e.g.~Lenzen \cite{lenzen:1987}), bipolar
outflows (see Bachiller \cite{bachiller:1996}) and disks (e.g.~McCaughrean
\& O'Dell \cite{caughodell:1996}). Even the progenitors of these YSOs,
starless dense cloud cores, seem to appear as elongated structures at
millimeter wavelengths (e.g.~Myers et al.~\cite{meyersfuller:1991}),
indicating that even in the early stages of star formation spherical
symmetry does not apply. The case of post-AGB stars and Planetary Nebulae is
just as compelling, with the majority of these nebulae being bipolar. The
Cygnus Egg Nebula (CRL 2688) and te Red Rectangle (HD 44179) are perhaps the
most spectacular examples of such bipolarity.

To model such objects, clearly one must resort to multi-dimensional transfer
computations. There is a vast literature on this topic. Methods roughly fall
in one of three catagories: Monte Carlo methods, Discrete Ordinate methods
and Angular Moment methods. Monte Carlo codes are very flexible and can be
used for a large variety of problems in multidimensional geometries, such as
UV continuum transfer (e.g.~Spaans \cite{spaans:1996}), optical and infrared
continuum transfer (Wolf et al.~\cite{wolfhenning:1999}), molecular line
transfer (Hogerheijde \cite{hogerheijdethesis:1998}), and Compton scattering
(e.g.~Pozdnyakov, Sobol \& Sunyaev 1977; Haardt \& Maraschi 1991). Such
methods perform well at low to medium optical depths, but it is well known
that at high optical depths they converge very slowly. 

Angular Moment methods, on the other hand, are very well suited to treat the
high optical depths regime, since they are related to (or variants of) the
diffusion equation (see e.g.~Yorke et al.~\cite{yorkebodlau:1993},
Sonnhalter et al.~\cite{sonnpreiyo:1995}, Murray et
al.~\cite{murraycas:1994}). However, it is not surprising that they fail at
low optical depth, since the diffusion approximation was never meant for
this regime.

In the Discrete Ordinate approach, not only space is discretized into cells,
but also the photon propagation direction. Most multi-dimensional
implementations of the Discrete Ordinate methods are based on the 
``Lambda Iteration'' scheme (e.g.~Collison \& Fix \cite{collfix:1991},
Efstathiou \& Rowan-Robinson \cite{efstrow:1991}, Philips
\cite{rphilips:1999}). The advantage of these methods over the Monte Carlo
approach is that they do not involve random noise, and therefore provide
`cleaner' answers. But they suffer from the same convergence problems as
Monte Carlo methods. However, for Lambda Iteration there are various ways to
cure this disease. The most well known of these methods is Accelerated
Lambda Iteration (ALI, e.g.~Scharmer \cite{scharmer:1981}, Rybicki \& Hummer
\cite{rybhum:1991}).

In this paper we will focus on the Discrete Ordinate approach to radiative
transfer because of its versatility, accuracy and the wide range of
convergence acceleration techniques available. However, despite the relative
efficiency of these methods, multidimensional calculations remain costly.
Feasibility constraints can pose severe limits on the spatial and
angular resolution, which could easily result in unacceptable numerical
diffusion. Also, this limits the number of models one can reasonably make to
fit observations, which could lead to dangerous undersampling of the parameter
space.

The bottleneck lies in the integration of the formal transfer equation. The
most straightforward way of performing these integrals is by the method of
``Long Characteristics'', which is accurate, but rather costly in CPU time.
A more efficient algorithm for doing this in two dimensions is the method of
``Short Characteritics'' (SC; Mihalas et al.~\cite{mam:1978}, Kunasz \& Auer
\cite{kunauer:1988}, Auer \& Paletou \cite{aupal:1994}, Stone et
al.~(\cite{stonemihnor:1992}). These algorithms are designed specifically
with cartesian or cylindrical coordinates in mind, and are not
straightforward to generalize to other coordinate systems. For circumstellar
envelopes, however, there are several arguments favoring the use of
spherical (polar) coordinates, as opposed to cylindrical coordinates. Most
circumstellar nebulae have density and temperature profiles that are peaked
towards the center. This means that the radiation field is dominated by
photons emitted in the central regions, which are subsequently reprocessed
in the outer parts of the nebula. The numerical scheme must therefore be
able to resolve both the very concentrated central regions and the extended
outer regions simultaneously. Also, it must guarantee that all radiation
emitted at small radii will eventually emerge at large radii, which amounts
to saying that flux must be conserved over a large range of radii.  Using
spherical coordinates and a logarithmic radial grid is the most natural way
to cover such a large dynamic range and guarantee flux conservation.

The goal of this paper is to describe, test and demonstrate an algorithm
that generalizes the Short Characteristics method to spherical
coordinates\footnote{Just prior to submission we became aware of a paper by
Busche \& Hillier (\cite{buschehillier:2000}), who describe a method of
short characteristics in spherical coordinates that is quite similar to
ours.}.  It is an algorithm specifically suited for axisymmetric
circumstellar nebulae and disks. It has been implemented in a multi-purpose
radiative transfer code named RADICAL, which is designed to perform 2-D
computations in dust continuum emission/absorption, multiple scattering off
dust grains, non-LTE line transfer, and (for application to X-ray binaries
and AGN) Comptonization. In this paper we describe the method of short
characteristics in spherical coordinates, and focus our attention to the
cases of simple isotropic scattering off dust grains and non-LTE line
transfer for multi-level molecules. For a more extensive discussion of the
algorithm and its applications, see Dullemond (\cite{dullemondthesis:1999}).

The structure of this paper is as follows. In Section
\ref{sec-general-method} we will present the equations of transfer we wish
to solve. In Section \ref{sec-sc-cart} we shall review the method of short
characteristics as it is often presented in the literature. In Section
\ref{sec-sc-spher} we will show how this method can be generalized to
spherical coordinates. Then we will put the algorithm to the test in Section
\ref{sec-test-cases}. Finally we will present two example applications in
Sections \ref{sec-eggneb} and \ref{sec-hcb}.

\section{Equations of radiative transfer}\label{sec-general-method}
The method that will be described in this paper is an Accelerated Lambda
Iteration method. In such an algorithm the integration of the formal
transfer equation is performed using a ``Lambda Operator''. In this section
we will present the equations that are to be solved, and we define the
Lambda Operator. The numerical details of the Lambda Operator will be given
in the next sections.

The formal transfer equation is
\begin{equation}\label{eq-trans-general}
\frac{dI_{\nu}}{ds} = \alpha_{\nu} (S_{\nu} - I_{\nu})
\comma
\end{equation}
with $I_{\nu}$ is the intensity, $S_{\nu}$ the source function,
$\alpha_{\nu}$ the opacity, and $s$ the path length. This equation must hold
along every straight line through the medium. Its integral form along a ray
through a point $P$ reads:
\begin{equation}\label{eq-basic-integral}
I_{\nu}(P) = e^{-\tau_\nu}I_{\nu}(0) + \int_0^{\tau_\nu} e^{-\tau'_\nu}
S(\tau'_\nu)d\tau'_\nu
\comma
\end{equation}
where $\tau_\nu$ is the optical depth along the ray, between point $P$
and the edge of the medium. After evaluating this integral for all angles
$\solidang$, one can compute the mean intensity $J_\nu$
\begin{equation}
J_\nu = \frac{1}{4\pi} \int I_\nu(\solidang) d\solidang
\fullstop
\end{equation}
The entire operation of computing $J_\nu(P)$ at every point $P$, for a given
source function $S_\nu$, can be written as the action of a linear Lambda
Operator $\Lambda$:
\begin{equation}
J_\nu = \Lambda [S_\nu]
\comma
\end{equation}
Using this Lambda Operator we can write down the complete transfer equation
for a simple problem of thermal emission and isotropic (dust) scattering
\begin{equation}\label{eq-srcfnc-simple}
S_{\nu} = \epsilon B_\nu(T) + (1-\epsilon) \Lambda[S_\nu]
\comma
\end{equation}
where $\epsilon\equiv\alpha^{\therm}_\nu/\alpha_\nu$ is the thermalization
coefficient (with $\alpha^{\therm}_\nu$ the thermal absorption opacity), and
$B_\nu(T)$ is the Planck function. Solving the transfer problem for
isotropic scattering and thermal emission amounts to solving
Eq.(\ref{eq-srcfnc-simple}) for $S_\nu$. The Lambda Iteration procedure
amounts to iteratively applying the Lambda Operator and computing the new
$S_\nu$ until convergence is reached. The Accelerated Lambda Iteration
procedure, which converges much faster, is a variant of this procedure,
involving an approximate operator $\Lambda^{*}$. For details we refer to
Hubeny (\cite{hubeny:1989}) and Rutten (\cite{rutten:1999}).

For multi-level line transfer, we follow the treatment of Rybicki \& Hummer
(\cite{rybhum:1991}). Consider an atom or molecule having $N$ levels, with
spontaneous radiative downward transition rates $A_{ij}$, Einstein
coefficients $B_{ij}$ and collision rates $C_{ij}$ between levels $i$ and
$j$. The line profile function $\tilde \varphi_{ij}(\nu)$ determines at
which frequencies the line emits and absorbs. When no systematic fluid
velocities are present, the line profile function is isotropic, and is
normalized to unity. For the application to circumstellar envelopes, the
dominant broadening mechanisms are turbulent and thermal broadening.  These
two mechanisms produce a Gaussian profile:
\begin{equation}\label{eq-turb-lineprof-def}
\tilde\varphi_{ij}(\nu) = \frac{c}{\lwtot\nu_{ij}\sqrt{\pi}} 
\exp\left(-\frac{c^2(\nu-\nu_{ij})^2}{\lwtot^2\nu_{ij}^2}\right)
\fullstop
\end{equation}
Here $c$ is the speed of light, $\nu_{ij}$ the line-center frequency of
the transition between levels $i$ and $j$, and $\lwtot$ is the line width,
\begin{equation}
\lwtot = \turblw + \sqrt{\frac{2 k T_{\kin}}{m_{\molec}}}
\comma
\end{equation}
where $T_{\kin}$ is the (kinetic) temperature of the gas, $m_{\molec}$
the mass of the molecule, and $\turblw$ is the turbulent line width. A
systematic fluid velocity can cause the line profile function to be
angle-dependent in the lab frame as a result of Doppler shift,
\begin{equation}
\varphi_{ij}(\solidang{},\nu) = \tilde\varphi_{ij}\big(\nu(1-\solidang{}\cdot \vec v/c)-\nu_{ij}\big)
\comma
\end{equation}
The opacity in the line associated with this line profile is:
\begin{equation}
\alpha_{ij}(\solidang{},\nu) = \frac{h\nu}{4\pi}N(n_jB_{ji}-n_iB_{ij})
\varphi_{ij}(\solidang{},\nu) \label{eq-molec-extinct-def}
\comma
\end{equation}
where $n_i$ are the fractional level populations, and $N$ the number density
of molecules. We assume complete redistribution for the lines.
The source function is then independent of frequency and angle:
\begin{equation}
S_{ij} = \frac{n_iA_{ij}}{n_jB_{ji}-n_iB_{ij}}
\fullstop
\end{equation}
The transfer equation for this source function is then
\begin{equation}\label{eq-molec-rad-trans-eq}
\frac{dI_{ij}(\solidang{},\nu)}{ds} = 
\alpha_{ij}(\solidang{},\nu) \left[ S_{ij}- I_{ij}(\solidang{},\nu)\right]
\comma
\end{equation}
where we assume non-overlapping lines.

The source term $S_{ij}$ is known once the fractional level populations
$n_i$ are known. They are a solution of the statistical equilibrium
equation. Using the definition of the line-integrated Lambda Operator
$\bar\Lambda_{ij} [S_{ij}] \equiv \bar J_{ij}$, with
\begin{equation}\label{eq-line-av-mean-int}
\bar J_{ij}= \frac{1}{4\pi}\int I(\solidang{},\nu)
\varphi_{ij}(\solidang{},\nu) d\solidang{} d\nu
\comma
\end{equation}
the statistical equilibrium equations become:
\begin{equation}\label{eq-stat-equil-lambda-op}
\begin{split}
\sum_{j>l} & \big[n_jA_{jl}+(n_jB_{jl}-n_lB_{lj})\bar\Lambda_{jl}
[S_{jl}]\big] \\
- &\sum_{j<l}  \big[n_lA_{lj}+(n_lB_{lj}-n_jB_{jl})\bar\Lambda_{lj}
[S_{lj}]\big] \\
+ &\sum_{j}\big[n_jC_{jl}-n_lC_{lj}\big]=0
\fullstop
\end{split}
\end{equation}
The non-locality of radiative transfer is now hidden in the $\bar\Lambda_{jl}$
operator, so that Eq.(\ref{eq-stat-equil-lambda-op}) now represents the
complete (non-linear) set of equations for line transfer. Lambda Iteration
now proceeds by iteratively applying the $\bar\Lambda_{jl}$ operator and
solving the matrix equation represented by
Eq.(\ref{eq-stat-equil-lambda-op}). Accelerated Lambda Iteration proceeds
according to the MALI scheme of Rybicki \& Hummer (\cite{rybhum:1991}).

\section{Short characteristics in cartesian coordinates}\label{sec-sc-cart}
To carry out the Lambda Iteration or Accelerated Lambda Iteration procedure,
we need a numerical implementation of the Lambda Operator $\Lambda$. In
Cartesian coordinates, the formal transfer equation Eq.(\ref{eq-trans-general})
becomes
\begin{equation}\label{eq-trans-cartesian}
\frac{dI_{\nu}}{ds}\equiv \omega_{x}\frac{\partial
I_{\nu}}{\partial x} + \omega_{y}\frac{\partial I_{\nu}}{\partial
y} = \alpha_{\nu} (S_{\nu}-I_{\nu})
\end{equation}
where translational symmetry in the $z$--direction was assumed.

The numerical implementation of the Lambda Operator amounts to
integrating Eq.(\ref{eq-trans-cartesian}) for given
$S_{\nu}$ and $\alpha_{\nu}$. This must be done on a
2--dimensional spatial grid $\vec x=(x_i,y_j)$, for a discrete set
of directions $\solidang{}=\{\solidang_{i}\}$ and frequencies
$\nu=\{\nu_{i}\}$. This will provide the specific intensity
$I(x_i,y_j;\solidang{}_k,\nu_l)$ for all $i,j,k,l$. Let us focus
on a given point $P=(x_i,y_j)$ and on a single direction and
frequency, $\solidang{}=\solidang{}_k,\nu=\nu_l$. The integral of
Eq.(\ref{eq-trans-cartesian}) can be performed
numerically along the entire characteristic starting at the upstream
boundary, heading in the downstream direction (i.e.~the direction
where the radiation comes from) and ending at point $P$ (see
Fig.~\ref{fig-lc-cartesian}).
\ifthenelse{\figs=0}{}{
\begin{figure}
\mbox{} \centerline{\epsfxsize=7cm \epsfysize=5cm
\epsfbox{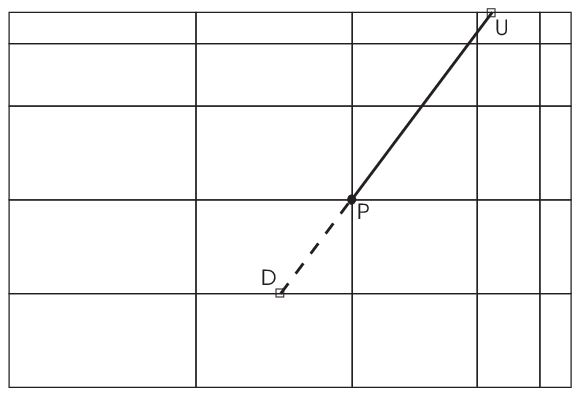}} \mbox{} \caption{An illustration of the Long
Characteristics (LC) method in Cartesian coordinates. The intensity at point
$P$ is computed integrating the transfer equation along the entire ray from
the upstream boundary (point $U$) towards point $P$. }
\label{fig-lc-cartesian}
\end{figure}}
This direct approach is called the method of Long Characteristics (LC).
Provided the discretization in angle $\solidang$ is appropriate, this method
is quite accurate and reliable. But it has a computational redundancy,
and hence it is overly time--consuming. Consider, for instance, a spatial
grid $\NN\times\NN$, a set of $\NN_{\solidang{}}$ directions and of
$\NN_{\nu}$ frequencies. The long characteristics integral of
Eq.(\ref{eq-trans-cartesian}) typically requires in the order of $\NN$
integration steps. This means that while the dimension of the 
grid is $\NN^2\times \NN_{\solidang{}}\times \NN_{\nu}$, the total
computational time scales as
\begin{equation}
t_{CPU} \propto \NN^3\times \NN_{\solidang{}}\times \NN_{\nu}\,.
\end{equation}

The Short Characteristics method of integration (SC; see
Fig.~\ref{fig-sc-cartesian}) does not have this redundancy.
\ifthenelse{\figs=0}{}{
\begin{figure}
\mbox{}
\centerline{
\epsfxsize=7cm
\epsfysize=5cm
\epsfbox{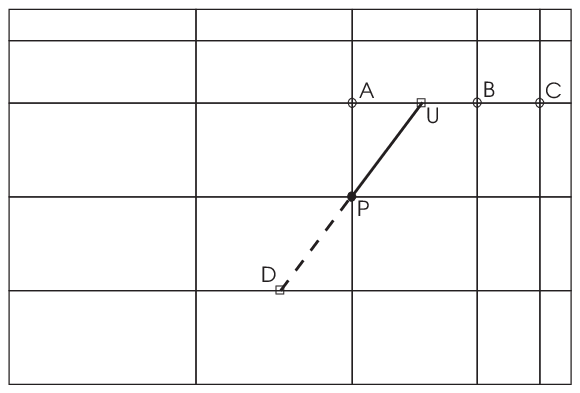}}
\mbox{}
\caption{An illustration of the SC method in
Cartesian coordinates. The short characteristic is the line connecting
point $U$ to $D$ through $P$. The value of the intensity at $U$ is
determined by quadratic interpolation
between the points $A$, $B$ and $C$.}
\label{fig-sc-cartesian}
\end{figure}}
Instead of performing the integral along the entire ray (the long
characteristic), we perform the integral only along that portion of the ray
(the short characteristic) which connects a point $U$ on the grid upstream of
$P$ to the closest intersection downstream of $P$ itself. The intensity at $P$
is given by
\begin{eqnarray}\label{eq-int-sc}
I_{\nu}(P;\solidang{}) &=& e^{-\tau_\nu}I_{\nu}(U;\solidang{})+\nonumber\\
& &\int_{0}^{\tau_\nu} e^{-\tau_{\nu}'}
S_{\nu}(\vec x(\tau_{\nu}');\solidang{})d\tau_{\nu}'
\end{eqnarray}
where $\tau_\nu$ is the optical depth between points $U$ and $P$.
The upstream intensity $I_{\nu}(U;\solidang{})$ can be found from
the intensities at $A$, $B$ and $C$ by 3--point quadratic
interpolation
\begin{equation}\label{interp-spat-I}
I_{\nu}(U;\solidang{}) = aI_{\nu}(A;\solidang{}) + bI_{\nu}(B;\solidang{}) +
cI_{\nu}(C;\solidang{})
\end{equation}
where $a$, $b$ and $c$ are the usual Lagrange coefficients for
polynomial interpolation. Quadratic or higher order
interpolation is necessary in order to reproduce the diffusion limit
for high optical depth, which is governed by a second order partial
differential equation.

The integral from $U$ to $P$ can be computed with second order accuracy by
interpolating the source function $S_{\nu}(\vec x(\tau_{\nu}');\solidang{})$
between the points $D$, $P$ and $U$. Following Olson \& Kunasz
(\cite{olsenkunasz:1987}), one finds
\begin{eqnarray}\label{eq-int-sc-threept}
I_{\nu}(P;\solidang{}) &=& e^{-\tau_{\nu}}I_{\nu}(U;\solidang{})
+u_{\nu} S_{\nu}(U;\solidang{}) \nonumber\\
& &\quad\quad+ p_{\nu} S_{\nu}(P;\solidang{}) + d_{\nu} S_{\nu}(D;\solidang{})
\comma
\end{eqnarray}
with
\begin{eqnarray}
u_{\nu} &=& e_0 + [e_2-(2\tau_{\nu}+\bar\tau_{\nu})e_1]/[\tau_{\nu}(\tau_{\nu}+\bar\tau_{\nu})]
\label{eq-a-value-sint}\\
p_{\nu} &=& [(\tau_{\nu}+\bar\tau_{\nu})e_1-e_2]/[\tau_{\nu}\bar\tau_{\nu}]\label{eq-b-value-sint}\\
d_{\nu} &=& [e_2-\tau_{\nu} e_1]/[\bar\tau_{\nu}(\tau_{\nu}+\bar\tau_{\nu})]\label{eq-c-value-sint}\\
e_0 &=& 1-e^{-\tau_{\nu}} \\
e_1 &=& \tau_{\nu} - e_0\\
e_2 &=& \tau_{\nu}^2 - 2e_1
\end{eqnarray}
where $\tau_{\nu}$ and $\bar\tau_{\nu}$ are the depths at $U$ and $D$,
respectively. It should be noted that this quadrature formula may have
pathological behaviour if the source function and/or the opacity varies
strongly between the points $U$, $P$ and $D$. This problem can be solved by
limiting the resulting integrals between zero and
$\max\big(j_{\nu}(P),j_{\nu}(U)\big)\,\Delta s$, where $\Delta s$ is the
path length along the short characteristic, and $j_\nu=\alpha_\nu S_\nu$.

By systematically performing the integrals over all the short
characteristics, one can find an approximate formal solution of
the transfer equation (Kunasz \& Auer \cite{kunauer:1988}, Auer \&
Paletou \cite{aupal:1994}, Auer et al.~\cite{auerfabtruj:1994}). A
key ingredient for the SC method to work is that the integrals
should be performed in the right order, so that the upstream
intensities $I_{\nu}(A;\solidang{})$, $I_{\nu}(B;\solidang{})$ and
$I_{\nu}(C;\solidang{})$ are known before the integral is
performed and Eq.(\ref{eq-int-sc-threept}) evaluated. In
order to do so, the grid must be swept from the two upstream
boundaries towards the two downstream boundaries.

The method of Short Characteristics is computationally less time
consuming than the method of Long Characteristics, 
because now the transfer integral is performed over a much shorter
path. For the same discretization introduced earlier in this
section, the computational time scales as
\begin{equation}
t_{CPU} \propto \NN^2\times \NN_{\solidang{}}\times \NN_{\nu}
\end{equation}
which is typically an factor $\NN$ shorter than in the case of Long
Characteristics.

\section{Short characteristics in spherical coordinates}\label{sec-sc-spher}
We now wish to generalize the Short Characteristics to spherical
coordinates. In the following we refer to a standard spherical coordinate
system $(R,\Theta,\Phi)$ where $\Theta$ is the latitude and $\Phi$ the
azimuth. By assuming axial symmetry, any dependence on $\Phi$ is suppressed,
although radiation is still allowed to travel along $\partial/\partial\Phi$,
as well as in the radial and meridional directions.

In order to describe the radiation field at each spatial point $P =
(R,\Theta)$ we need to set up a local coordinate system to characterize the
photon direction at $P$. We introduce two independent angles on the sky of
the local observer: $\theta$ and $\phi$. The north pole of this local
sky-map is chosen to coincide with the outward--pointing radial
direction. The $\phi$ angle is gauged in such a way that $\phi=0$ points
parallel to the equator of the global coordinate system (see
Fig.~\ref{fig-setup-angcoord}).
\ifthenelse{\figs=0}{}{
\begin{figure}
\mbox{} \centerline{\epsfxsize=6cm \epsfysize=7cm
\epsfbox{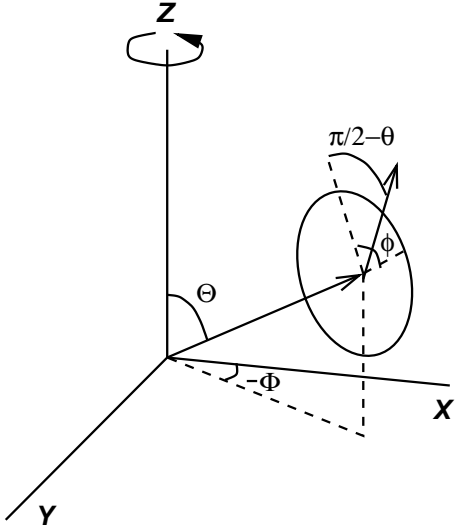}} \mbox{} \caption{The global and
local coordinate systems used to describe the radiation field.
Here $\pi/2-\theta$ is shown  instead of $\theta$ for the clarity
of illustration.} \label{fig-setup-angcoord}
\end{figure}}
As is customary in transfer theory, we use $\mu\equiv \cos\theta$ instead of
$\theta$ itself, so the specific intensity depends upon the two spatial
variables $R,\Theta$, the ray direction $\mu,\phi$, and the frequency $\nu$,
$I = I_{\nu}(R,\Theta\,;\,\mu,\phi)$.  The transfer equation,
Eq.(\ref{eq-trans-general}), in spherical coordinates reads
\begin{equation}
\begin{split}\label{eq-formal-trans-equation}
\frac{d I_{\nu}}{ds} \equiv \mu\frac{\partial I_{\nu}}{\partial R}
& - \frac{\sqrt{1-\mu^2}}{R}\sin\phi\frac{\partial I_{\nu}}{\partial\Theta}
+ \frac{1-\mu^2}{R}\frac{\partial I_{\nu}}{\partial\mu} \\
&-\frac{\cos\phi}{\tan\Theta}\frac{\sqrt{1-\mu^2}}{R}
\frac{\partial I_{\nu}}{\partial\phi} = \alpha_{\nu}(S_{\nu}-I_{\nu})
\fullstop
\end{split}
\end{equation}

An important consequence of the use of spherical coordinates is that,
contrary to what happens for cartesian coordinates, the photon angles $\mu$
and $\phi$ are no longer constant along the rays. The variation of $R$,
$\Theta$, $\mu$ and $\phi$ along the path are
\begin{xalignat}{2}
\frac{dR}{ds} &= \mu\,, &
\frac{d\Theta}{ds} &= -\frac{\sqrt{1-\mu^2}}{R}\sin\phi\,,
\label{eq-geodesic-rtheta}
\\
\frac{d\mu}{ds} &= \frac{1-\mu^2}{R}\,, &
\frac{d\phi}{ds} &=
-\frac{\sqrt{1-\mu^2}}{R}\frac{\cos\phi}{\tan\Theta}\,,
\label{eq-geodesic-muphi}
\end{xalignat}
where $s$ is the path length. Solving these equations yields
\begin{xalignat}{1}
R^2 &= \bimpact^2 + \saffine^2 \comma\label{eq-char-sol-r}  \\
\cos\Theta &= \frac{z_0 + \saffine\,\cos\thinf}{\sqrt{\bimpact^2 + \saffine^2}}
\comma\label{eq-char-sol-theta}  \\
\mu &= \frac{\saffine}{\sqrt{\bimpact^2 + \saffine^2}}\comma \label{eq-char-sol-mu}\\
\sin\phi &= \frac{\bimpact^2\cos\thinf-z_0\saffine}
{\bimpact\sqrt{\bimpact^2+\saffine^2-
(z_0+\saffine\cos\thinf)^2}} \label{eq-char-sol-phi}
\comma
\end{xalignat}
where $b$ is the impact parameter of the ray with respect to the origin,
$z_0$ is the height above the midplane of closest approach to the
symmetry-axis, and $\Theta_0$ is the inclination at infinity. When
projected into the subspace spanned by $R,\Theta$ the trajectory becomes
a hyperbola, as is shown in Fig.~\ref{fig-long-charac}. We stress that
this shape is caused by eliminating the dependence on the $\Phi$--angle, and
is purely a projection effect.
\ifthenelse{\figs=0}{}{
\begin{figure}
\mbox{}
\centerline{\epsfxsize=8cm
\epsfysize=8cm
\epsfbox{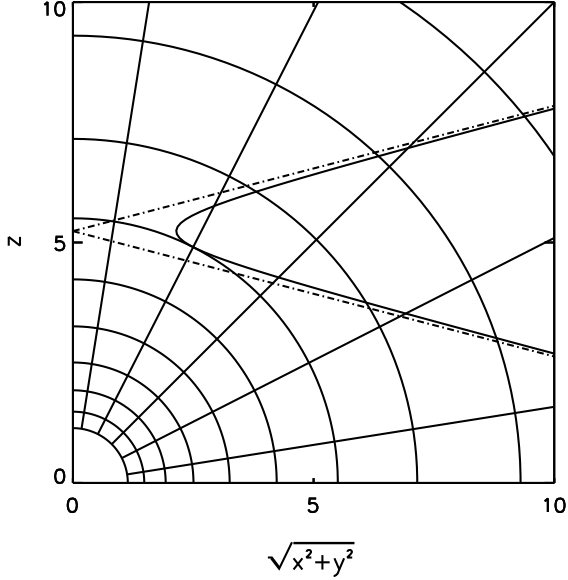}}
\mbox{}
\caption{An example of a long characteristic in an axially symmetric
space. Only the upper quadrant is shown. The vertical axis is the symmetry
axis and the horizontal axis the equator. The bot-dashed lines represent the
asymptotes of the hyperbolic characteristic.}\label{fig-long-charac}
\end{figure}}

For the numerical implementation of the short characteristics
scheme, we are interested in those characteristics that pass
through a grid point $P=(R_k,\Theta_l)$ and are tangent to one of
the local discrete ordinates $(\mu_i,\phi_j)$. Clearly, once
$R_k,\Theta_l,\mu_i,\phi_j$ are fixed, such a characteristic is
unique and the values of its parameters are
\begin{xalignat}{2}\label{eq-sol-sc-angle}
\bimpact^2 &= R_k^2(1-\mu^2_i) \\
\cos\thinf &= \mu_i\cos\Theta_l + \sqrt{1-\mu_i^2}\,\sin\Theta_l
\sin\phi_j\\
z_0 &= R_k \big[ (1-\mu^2_i)\,\cos\Theta_l - \nonumber\\
&\quad\quad\mu_i\sqrt{1-\mu^2_i}
\sin\Theta_l\sin\phi_j \big]\,.
\end{xalignat}
The short characteristic passing through
$(R_k,\Theta_l\,;\,\mu_i,\phi_j)$ is defined as the section of
this curve that starts at the closest intersection with the grid
lines {\em upstream} of $P$ (point $U$), passes through $P$ and ends
at the closest intersection with the grid lines {\em downstream} of
$P$ (point $D$). The location of the points $U$ and $D$ is
specified by the corresponding values of parameter $s$ along the
ray, $s_U$ and $s_D$, which are found solving equations
(\ref{eq-char-sol-r})--(\ref{eq-char-sol-theta}) with $R=R_K$ and
$\Theta=\Theta_L$, where $K = k-1, k, k+1$ and $L = l-1, l, l+1$.
Both $R=R_k$ and $\Theta=\Theta_l$ need to be included because the
characteristic may intersect the same $\Theta$ or $R$ grid line
twice. In principle, each equation has two solutions for a given
value of $K$ and $L$, giving 12 possible roots
\begin{eqnarray}
s_{1\ldots 6} &=& \pm \sqrt{R_K^2-b^2} \label{rootr}\\ s_{7\ldots
12} &=& \frac{1}{\cos^2\Theta_{\infty}-\cos^2\Theta_L}
 \bigg\{-z_0\cos\Theta_{\infty}\pm \nonumber\\
& & \cos\Theta_L
\sqrt{b^2(\cos^2\Theta_{\infty}-\cos^2\Theta_L)+z_0^2}\bigg\}\,.
\end{eqnarray}
However, two of these solutions always give $s=s_P$,
i.e.~$P=(R_k,\Theta_l)$ itself, and are of no interest. Of the remaining
10 roots, some are complex and must be rejected.
Between the real solutions, the one representing point $D$ ($U$)
is selected asking that  $s>s_P$ ($s<s_P$) and that $|s-s_P|$ is minimum.
For convenience, in the following we will denote with
$\tilde R_{1}$, $\tilde\Theta_{1}$, $\tilde\mu_{1}$
and $\tilde\phi_{1}$ the values of the independent variables along the ray
at point $U$.

Although, as we have just shown, short characteristics can be easily defined
in spherical coordinates, two major problems have to be solved before they
can be of any use in building a transfer algorithm. The first point concerns
the fact that, as it was mentioned earlier, $\mu$ and $\phi$ change along
the ray. This means that in addition to spatial interpolation (see
Eq.~\ref{interp-spat-I}), we are forced to interpolate in $\mu$ and $\phi$
as well in order to evaluate $I_\nu(U,\solidang{})$. This is because the
intensities at points $A$, $B$ and $C$ are known only for a discrete set of
directions which are different, in general, from
($\tilde\mu_1\,,\tilde\phi_1)$.

The second, more fundamental difficulty arises because in spherical
coordinates the concept of upstream and downstream boundaries is different
from the Cartesian case. Radial infinity is both the upstream and the
downstream boundary, while in $\Theta$ there is no obvious upstream or
downstream boundary. If the grid is swept from $\Theta=0$ to $\Theta=\pi$ or
vice versa, one will encounter situations in which the intensity at one of
the points $A$, $B$, $C$ is not known before the evaluation the transfer
integral along the short characteristic (Eq.~\ref{eq-int-sc}) is
performed. An example of such a situation is shown in
Fig.~\ref{fig-sc-spher-theta-extr}. Interpolation makes use of the points
$A$, $B$ and $C$, but since $A$ coincides with $P$, the intensity at the
point $A\equiv P$, has not been computed yet.  \ifthenelse{\figs=0}{}{
\begin{figure}
\mbox{} \centerline{\epsfxsize=8cm \epsfysize=8cm
\epsfbox{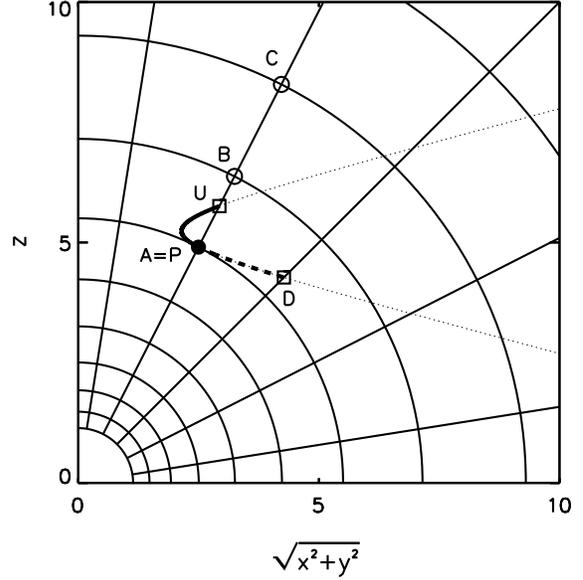}} \mbox{} \caption{An example
of a Short Characteristic (SC) in spherical coordinates (the heavy
line connecting $U$ to $D$ through $P$). As in Fig.~\ref{fig-sc-cartesian} the downstream part of the SC (which is only
needed to guarantee second order accuracy in the formal integral
between $U$ and $P$) is dashed. The dotted line shows the complete
ray to which the SC belongs.
} \label{fig-sc-spher-theta-extr}
\end{figure}}

\subsection{Extended Short Characteristics}
The problem of unknown intensities can be solved by modifying the definition
of short characteristics to be the part of the ray that connects $P$, not
with just the nearest gridline intersection, but with the nearest $R=R_k$
gridline intersection, i.e.~the nearest radial shell. Such an ``extended
short characteristic'' (ESC) is illustrated in Fig.~\ref{fig-local-char}.
\ifthenelse{\figs=0}{}{
\begin{figure}
\mbox{} \centerline{\epsfxsize=8cm \epsfysize=8cm
\epsfbox{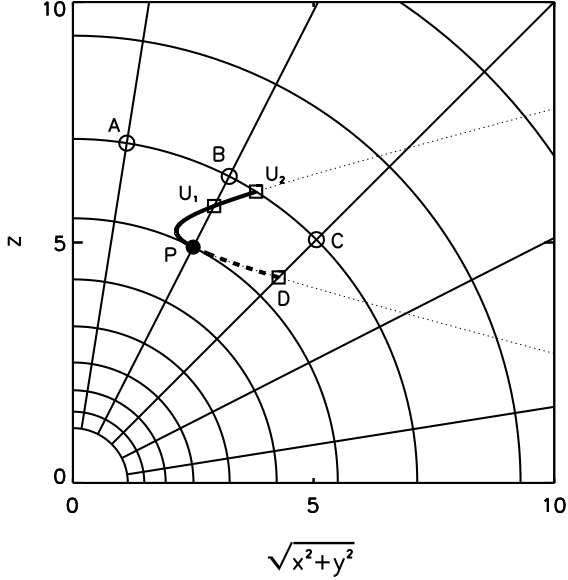}} \mbox{} \caption{ An example of
an Extended Short Characteristic (ESC) in spherical coordinates
(the heavy line connecting $U_2$ to $D$ through $P$). It is the
extended version of the SC shown in Fig.~\ref{fig-sc-spher-theta-extr}. The ESC does not have the same
problems as the SC because for the ESC none of the points $A$, $B$
or $C$ coincides with point $P$.
}
\label{fig-local-char}
\end{figure}}
The starting point $U$ of such an ESC will be located either at $R=R_{k-1}$,
$R=R_{k+1}$ or back at $R=R_k$. This means that in between $U$ and $P$ the
ESC may intersect one or more $\Theta$ grid lines. The point $D$ on the
downstream side remains the same as for standard SCs.

By using ESC instead of SC the ``problem of unknown upstream intensities''
can be eliminated. In fact, if a proper sweeping scheme is chosen (see
Subsection \ref{subsec-sweeping}), the problem of unknown intensities only
occurs in those situations when a short characteristic curves back onto the
same $\Theta$-gridline from which it originates, as is illustrated in
Fig.~\ref{fig-sc-spher-theta-extr}. By extending only those short
characteristics, and leaving the rest truly short, one can also avoid
unknown intensities in the sweeping scheme. Just for notation we call this
scheme the Minimally Extended Short Characteristics scheme (MESC). MESC is
almost as accurate as ESC, but ESC is more closely similar to its
one-dimensional spherical analogues, and is slightly less numerically diffusive.

In the following we denote with $D$ (or $U_{-1}$) the single downstream
intersection with a grid line, and with $U_i \ (i=1,\ldots, m)$ the multiple
intersections upstream of $P$. The point $U_m$ is therefore the true
upstream starting point of the ESC, where the intensity must be found by
interpolation. In Fig.~\ref{fig-local-char} this is the point $U_2$ and
the ESC consists of two segments in this case.

\subsection{The sweeping scheme}\label{subsec-sweeping}
Using the (minimally) extended short characteristics defined above, we can
systematically sweep the grid without encountering unknown intensities. We
start at the outer boundary $R_\infty$ and integrate inwards only those ESCs
for which $ \mu_i\le 0$. The sweeping order in $\Theta$ is from $\Theta=0$
to $\Theta=\pi$ and then back.

The intensity at each $P=(R_k,\Theta_l)$ is found for all $\mu_i\le 0$ and
$\phi_j$ by tracing the ESCs back to their upstream starting point $U_m$.
At point $U_m$ the values of $R$, $\Theta$, $\mu$ and $\phi$ are different
from those at $P$, and will be denoted with $\tilde R_m$, $\tilde\Theta_m$,
$\tilde\mu_m$ and $\tilde\phi_m$. The find the intensity at $(\tilde
R_m,\tilde\Theta_m,\tilde\mu_m, \tilde\phi_m))$ by interpolation.  We then
integrate the formal transfer equation along each segment of the ESC
connecting $U_m$ to $P$, according to Eq.(\ref{eq-int-sc-threept}). This
gives
\begin{equation}\label{eq-int-sc-threept-spher}
\begin{split}
I_{\nu}(P) = & \exp(-\tau_m)I_{\nu}(U_m) +
\sum_{i=1,m}\exp(-\tau_{i-1})\times\\
&\big(u_{\nu,i}S_{\nu,i}+p_{\nu,i-1} S_{\nu,i-1}+d_{\nu,i-2}S_{\nu,i-2}\big)
\end{split}
\end{equation}
where $\tau_i$ is the depth from $P$ to $U_i$ and the index $i$
denotes the quantity evaluated at $U_i$ (e.g. $S_i\equiv S(U_i)$;
$i=0$ refers to $P$ and $i=-1$ to $D$). The $u$'s, $p$'s and $d$'s are
defined as in equations (\ref{eq-a-value-sint}),
(\ref{eq-b-value-sint}) and (\ref{eq-c-value-sint}), but with
$\tau$ replaced by $(\tau_{i}-\tau_{i-1})$ and $\bar\tau$ by
$(\tau_{i-1}-\tau_{i-2})$.

The integration is then repeated moving towards smaller radii, until the
inner boundary at $R=R_1$ is reached. Here we can include the contribution
of a central source or any other boundary condition.

Then we start integrating back towards larger radii, until we reach the
outer edge. By now the radiation field on the grid
$I_\nu(R_k,\Theta_l\,;\,\mu_i,\phi_j)$ is known.

\subsection{Tangent-ray discretization of photon direction}\label{sec-photon-sampling}
Equations (\ref{eq-char-sol-mu}) and (\ref{eq-char-sol-phi}) show that $\mu$
and $\phi$ change along the ESC. If we follow the ESC upstream towards a
point $U_m$ (see Fig.~\ref{fig-local-char}), then the values of these two
angles at $U_m$ are generally not exactly at the discrete values $\{\mu_i\}$
and $\{\phi_j\}$ of the sample of directions. This means that we must
interpolate not only in space (between $A$, $B$ and $C$), but also in
direction $\mu$, $\phi$. Although, this is not, in principle, a fundamental
problem, the use of interpolations should be reduced to a minimum to avoid
unnecessary numerical diffusion.

Fortunately one can eliminate the interpolation in $\mu$ by means of a
suitable choice of the $\mu$--grid so that all ESCs always start and end at
one of the $\{\mu_i\}$ points. We let the $\mu$--discretization depend on
$R_k$,
\begin{equation}
\{\mu_{k,j}\,;\,j=-m_k,\cdots,m_k\}
\comma
\end{equation}
and choose the $\mu_{k,j}$ in such a way that for each $j=-m_k,\cdots,m_k$
there is a $j'=-m_{k-1},\cdots,m_{k-1}$ such that
\begin{equation}\label{eq-mu-discr}
(1 - \mu_{k,j}^2)R^2_k = (1-\mu_{k-1,j'}^2)R^2_{k-1}
\fullstop
\end{equation}
This choice is based on the fact that the values of $\mu$ and $R$
along an ESC depend only on each other and on $\bimpact{}$, as 
can be seen by combining equations (\ref{eq-char-sol-r}) and
(\ref{eq-char-sol-mu}):
\begin{equation}\label{muandr}
(1-\mu^2)R^2=\bimpact{}^2
\fullstop
\end{equation}
By choosing the $\mu_{k,j}$ according to Eq.(\ref{eq-mu-discr}), a ray
which originates at $R_{k}$ with, say, $\mu=\mu_{k,j}$ and arbitrary $\phi$,
reaches any other radius along the path with a value of $\mu$ which
coincides with one of the points of the local $\mu$ grid there, thus
eliminating the need for interpolation.

A $\mu$-grid that is consistent with Eq.(\ref{eq-mu-discr}) is:
\begin{equation}\label{eq-mu-grid-tangent-ray}
\mu_{k,\pm i} = \pm \sqrt{1-\frac{R^2_{k-i}}{R^2_k}}
\comma
\end{equation}
in agreement with the angular spacing induced in spherical symmetry by the
``tangent ray method'' (see e.g.~Mihalas \cite{mihkunhum:1975}; Zane et
al. \cite{zanetur:1996}). Actually, it can be easily shown that in 1-D
spherical symmetry the ESC method, with $\mu_{k,j}$ given by
Eq.(\ref{eq-mu-grid-tangent-ray}), is fully equivalent to the tangent ray
method. This is an important feature of the algorithm since it then exactly
recognizes spherical symmetry. And, even in the absence of spherical
symmetry, it transports radiation outward without any numerical diffusion in
$R$ or $\mu$.

However, the $\mu$ spacing implied by Eq.(\ref{eq-mu-grid-tangent-ray}) has
the tendency to give a poor sampling around $\mu=0$. This problem can be
easily solved by introducing some (typically one or two) extra points around
$\mu\simeq 0$ to enhance the angular resolution there. Obviously this
violates the original prescription and therefore requires the use of
interpolation for these extra $\mu$--points, producing a small amount of
angular diffusion for $\mu\sim 0$. Generally this diffusion is small.

Unfortunately the interpolation in $\phi$ can never be avoided.  The $\phi$
angle depends in a complicated way on $s$ (see Eq.~\ref{eq-char-sol-phi})
and it can change rapidly even within one element of a ESC. Both first and
higher order interpolation in $\phi$ have been tested in our numerical
code. We have found that in most cases the $\phi$ diffusion is not very
large and, in general, influences the solution less than the spatial
($\Theta$) diffusion.

\subsection{Special treatment of radiation near $\mu\simeq 1$}
The tangent-ray discretization of $\mu$ allows the algorithm to accurately
conserve radial flux. However, such a choice of $\mu$--angles requires a
large number of $\mu$ points at larger radii, typically $m_k\gtrsim k$.  One
cannot make do with a smaller number of $\mu$ points without facing the risk
of loosing flux. This is illustrated in the following argument. If radiation
is emitted at a radius $R$, an observer at $R_k\gg R$ can see the radiation
from the emitting region even if its eyes cannot resolve the source. This is
because the observer's eyes measure the flux and not the intensity. The ESC
algorithm, on the other hand, deals with intensity, and intensity is
converted into flux by performing an integral over $d\mu\, d\phi$. For this
integral to be reasonably accurate, the emitting region must be resolved in
$\phi$ and $\mu$, leading to the requirement $m_k\gtrsim k$.

Unfortunately this means that the computational cost scales as $N_R^2$ if
one wishes to extend the span of the radial domain. Since the ability to
deal with many orders of magnitude in radius is crucial to solving transfer
problems in circumstellar envelopes, this scaling is undesirable.

An easy way to solve this scaling problem is related to the simple
observation that all photons with $\mu\simeq 1$ follow roughly a radially
outgoing trajectory and they tend to travel more radially (i.e.~with $\mu$
closer to unity), the further they propagate outwards. In the ``radial
streaming'' limit ($1-\mu\ll 1$), Eq.(\ref{eq-mu-discr}) becomes
approximately
\begin{equation}\label{eq-mu-farfield}
\frac{1 -\mu_{k,j}}{1-\mu_{k-l,j-l}}=
\frac{\Omega_{k,j}}{\Omega_{k-l,j-l}}\simeq
\frac{R^2_{k-l}}{R^2_k}
\end{equation}
where $\Omega_{k,j}$ is the solid angle (bounded by $\mu_{k,j}$)
at $R_k$. Eq.(\ref{eq-mu-farfield}) is just a restatement of
the $1/R^2$ law which is exactly obeyed by a point source and by
any radiation in the radial streaming limit.

This property of radially outward radiation makes it possible to bundle all
$\mu$--points with sufficiently large $\mu$ into a single collective
flux-like bin. The intensity of that bin will be treated as the {\em average
intensity} within that collective bin. The idea is to divide the
$\mu$--range $[-1,1]$ into three parts
\begin{xalignat}{3}
& [1,-\mu_{rs}] & & \hbox{inward intensity bin:} &  \mu &\simeq -1
\nonumber \\ & [-\mu_{rs},\mu_{rs}]& & \hbox{intensity samples:} &
\mu &= \mu_{k,j} \nonumber \\ & [\mu_{rs},1] & & \hbox{outward
averaged bin:} &  \mu &\simeq  1  \nonumber
\fullstop
\end{xalignat}
This way the number of $\mu$--gridpoints $m_k$ at each radius can be
limited, depending on how close $\mu_{rs}$ is to unity. We choose a global
value for $\mu_{rs}$, and do not allow this to differ from one radius to
another.

Because the radial outward bin represents an {\em integrated} intensity,
i.e.~an average of the true intensity over a solid angle
$\Omega_{rs}=2\pi(1-\mu_{rs})$, it requires a special treatment. Let us
denote the average intensity in this bin as $I^{+}_\nu(R_k,\Theta_l)$.  The
integration formula, Eq.(\ref{eq-int-sc-threept-spher}), for
$I^{+}_\nu(R_k,\Theta_l)$ becomes
\begin{equation}\label{eq-int-fluxbin}
\begin{split}
I^{+}_{\nu}(&R_k,\Theta_l) =
  \exp(-\tau)\bigg[\frac{R^2_{k-1}}{R^2_k}I^{+}_{\nu}(R_{k-1},\Theta_l) + \\
& \left(1-\frac{R^2_{k-1}}{R^2_k} \right)\frac{1}{2\pi}\int_0^{2\pi}
   I_{\nu}(R_{k-1},\Theta\,;\,\mu_{m_k-1},\phi)\,d\phi\bigg] \\
&+u_{\nu}S_{\nu}(R_{k-1},\Theta_l)+p_{\nu}S_{\nu}(R_k,\Theta_l) \\
&+d_{\nu}S_{\nu}(R_{k+1},\Theta_l)
\fullstop
\end{split}
\end{equation}
This formula simulates the correct $1/R^2$ behavior of the flux for
optically thin media provided $\mu_{rs}$ is sufficiently close to unity.  It
reduces to the standard expression when the medium is optically thick. An
estimate of the error introduced by the assumption of radial streaming can
be made by comparing equations (\ref{eq-mu-discr}) and
(\ref{eq-mu-farfield}), and is of order $(1-\mu_{rs})/2$. The radiation
contained in the $\mu=1$ bin will not accumulate any numerical interpolation
errors because it moves strictly along the radial grid lines.

The inward collective bin will always behave as a real intensity, so that
the $1/R^2$ behavior does not need to be simulated.

\subsection{Spectra and images}\label{subsec-spectra}
Once the iterative part of the transfer has been completed and the source
function is known, the next step is to produce images and spectra. An image
is produced by formal integration of the source function along long
characteristics through the medium (ray tracing). Each ray represents one
pixel of the image. One can produce spectra by making images at a range of
frequencies, and integrating these images over the ``detector'' aperture.

Here, as in the Lambda iteration, we face resolution problems if the source
under consideration spans a large range in $\log(R)$.  The central parts of
the image are often much brighter than the rest, but cover a much smaller
fraction of the image. The spectrum may therefore contain significant
contributions of flux from both the central parts and the outer regions of
the image. Unless the image resolve all spatial scales of the object, the
spectra produced in such a way are unreliable.

If a rectangular arrangement of pixels is used, one must make sure to use a
variable spacing in both $x$ and $y$, in such a way that the small scales
around the star are sufficiently resolved. If one is mainly interested in 
the images themselves, this seems the most reliable and straightforward
way to go.

For the production of spectra we propose a different approach. Rather than
arranging the pixels over a rectangle, we arrange them in concentric rings.
The impact parameters of the circles are related to the radial
grid points of the transfer calculation. For a reliable evaluation of the
spectra it is generally enough to have one circle for each $R_i$, plus some
more, about 5, to resolve the central region. The number of circles $N_b$ in
each image is therefore roughly the same as the number of radial grid
points: $N_b=N_R+5$. The number of pixels in each circle $N_\varphi$ is
slightly less straightforward to choose, but for reliable spectra it is
generally sufficient to take $N_\varphi=2N_\Theta$, where $N_\Theta$ is the
number of $\Theta$ grid points, counted from pole to pole. Using this
method, the images automatically resolve all relevant scales, while using
only a fairly limited number of pixels.

\section{Testing the ESC Lambda Operator}\label{sec-test-cases}
The Extended Short Characteristic implementation of the Lambda
Operator is not exact, as opposed to the one based on Long
Characteristics. The interpolations used in the ESC algorithm
introduce numerical diffusion, even in the optically thin regime, and
this constitutes a potential threat to the reliability of the
method. In order to test the accuracy of the ESC Lambda Operator we
have performed a series of runs for a number of simple setups, comparing
the results of the ESC calculation with those obtained by means of an
exact LC Lambda Operator. Here we present the analysis for three such
tests.
\subsection{Optically thick annulus}
The first test problem concerns the determination of the radiation
field produced by an
optically thick, isothermal, sharply--edged annuls, bounded by
$R_0<R<R_M$ and $\Theta_M<\Theta <\pi -\Theta_M$. In the actual
calculation we have taken $R_{0}= 2.86$, $ R_{M} = 4.83$,
$\Theta_M = 1.26$ and the absorbtion $\alpha_a$ is given by
\begin{equation}\label{eq-annulus-def}
\alpha_a(R,\Theta) = \left\{\begin{matrix}
10^3 & \hbox{inside the annulus}\\
0    & \hbox{elsewhere}\,. \\
\end{matrix}\right.
\end{equation}
For the sake of simplicity all variables are dimensionless and the
temperature has been taken
such that $B_\nu(T)=1$. We use a spatial grid with
$20$ radial points, logarithmically spaced such that
\begin{equation}
\frac{R_{i+1}-R_i}{R_i} = 1.1402\,, \nonumber
\end{equation}
and the angular grid in $\Theta$ consists of 20 equally spaced points from
pole to pole. Fig.~\ref{fig-thick-annul-overview}
\ifthenelse{\figs=0}{}{
\begin{figure}
\mbox{}
\hfil\epsfxsize=8cm
\epsfysize=8cm
\epsfbox{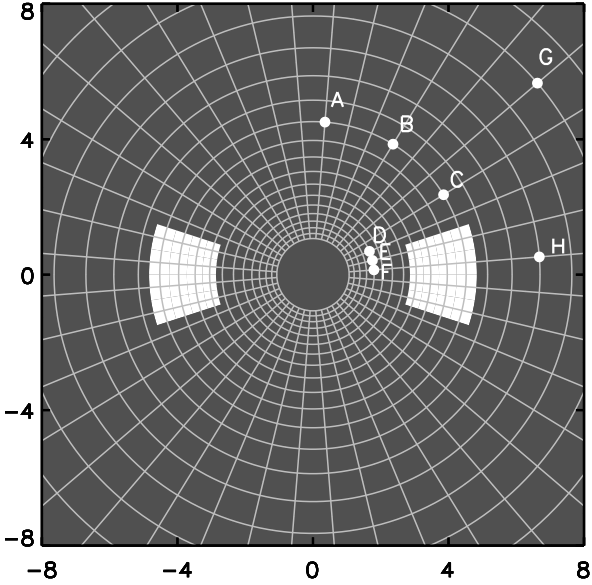}
\mbox{}
\caption{The set--up for the test runs. The annulus is
shown together with the spatial mesh. Here all four quadrants are shown,
although only the first quadrant needs to be computed as a result of
mirror symmetry
in the equatorial plane and cyclindrical symmetry around the polar axis.
Nine representative grid points, labeled $A-H$, are marked (see
Fig.~\ref{fig-thick-annul-radfield}).}
\label{fig-thick-annul-overview}
\end{figure}}
shows the configuration for the test problem. Mirror symmetry in
the equator reduces the number of actual points to 10 in the range
$0<\Theta <\pi/2$.  The mesh in photon momentum space consists of 32
equally spaced points in $\phi$, covering the range 0--$2\pi$, and 41
points in $\mu$, 38 chosen according to Eq.(\ref{eq-mu-discr}),
plus $\mu=0$ and $\mu=\pm 0.24$ to ensure sufficient resolution at
small $\mu$. Because of the symmetry, the transfer needs to be solved
only for the 16 points in the ranges $0<\phi\le \pi/2$ and
$3\pi/2<\phi\le 2\pi$.

The radiation field emitted by such a source can be
semi--analytically determined. As seen by an observer at some point $P$,
it is simply the projection of the object on the sky of the observer.
Since
the object is sharply--edged and highly optically thick ($\tau\gg
100$), its projection will be sharply--edged as well. The intensity is
simply given by
\begin{equation}\label{eq-projection-on-sky}
I_\nu(P\,;\,\mu,\phi) = \left\{\begin{matrix}
0 & \hbox{for rays missing the object} \\
B_{\nu}(T) & \hbox{for rays hitting the object} \\
\end{matrix}\right.\;,
\end{equation}
so it is easy to compute the projections on the sky of the observer at
various positions in space by using some independent ray--tracing
algorithm or a semi--analytical computation of the image. This image
can then be compared to that produced by the ESC and the LC transfer
algorithms to evaluate the accuracy of the ESC Lambda Operator.

Let $I_{ESC}(\mu,\phi)$ and $I_{LC}(\mu,\phi)$ be the intensities, at
the observer location, computed using the ESC and the LC algorithms,
for the same discretization in $\mu$ and $\phi$. Let, moreover,
$I(\mu,\phi)$ be the true intensity, which may be found by tracing
individual rays with very high resolution in $\mu$ and $\phi$. We
define the standard error of the ESC algorithm as
\begin{equation}
\sigma^2_{ESC} = \frac{\int (I_{ESC}-I)^{2} d\mu d\phi}
{\int I^{2} d\mu d\phi}
\end{equation}
and the error in the mean intensity $J$ as
\begin{equation}
\epsilon_{ESC} = \frac{\int (I_{ESC}-I) d\mu d\phi}
{\int I d\mu d\phi} \equiv \frac{J_{ESC}-J}{J}\,.
\end{equation}
Similar definitions apply to the LC errors.

Radiative transfer for this setup has been performed using the ESC
method. The results are shown in Fig.~\ref{fig-thick-annul-radfield} for the
9 gridpoints labeled $A-I$ in Fig.~\ref{fig-thick-annul-overview}.
\ifthenelse{\figs=0}{}{
\begin{figure*}
\mbox{}
\centerline{
\epsfxsize=4cm\epsfysize=4cm\epsfbox{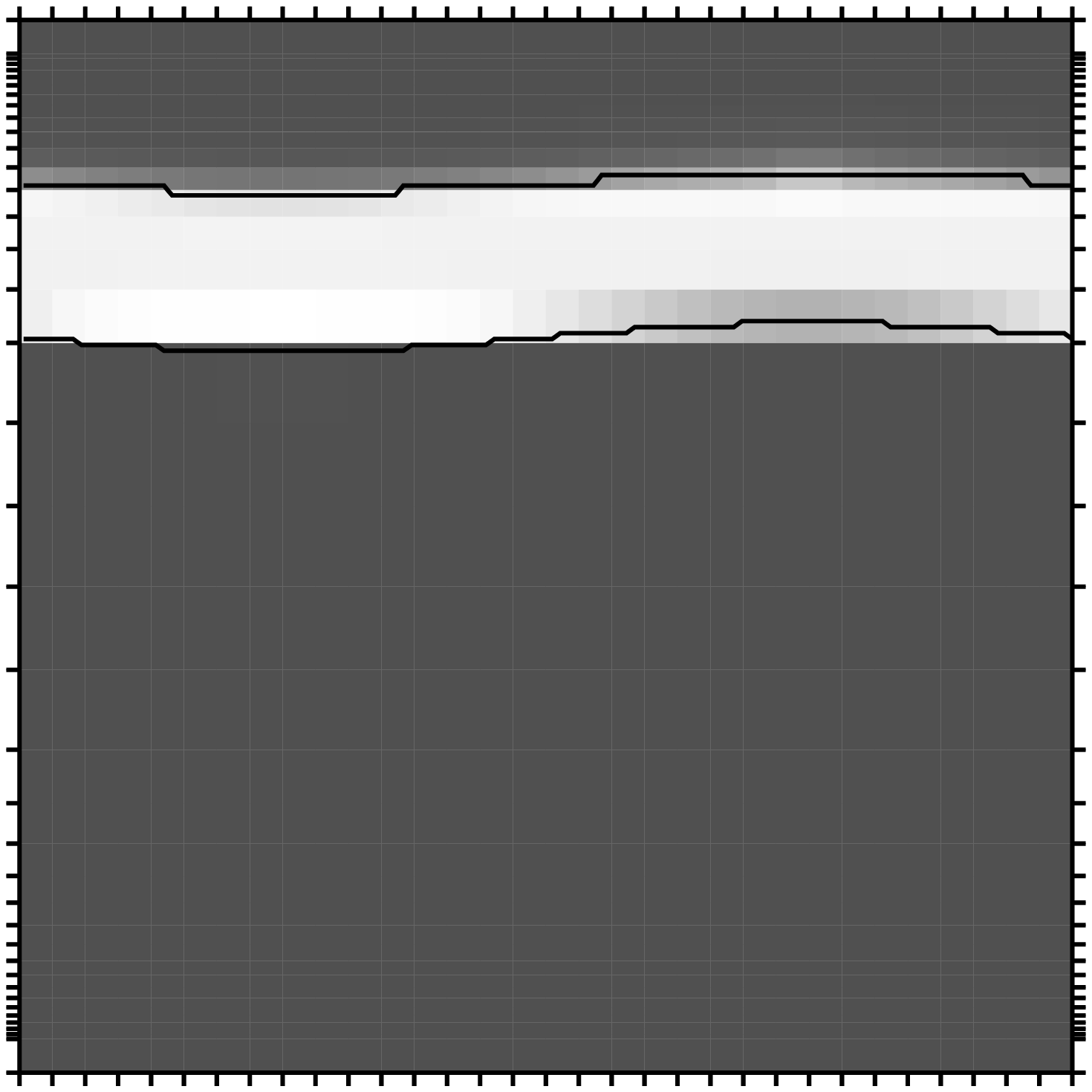}
\epsfxsize=4cm\epsfysize=4cm\epsfbox{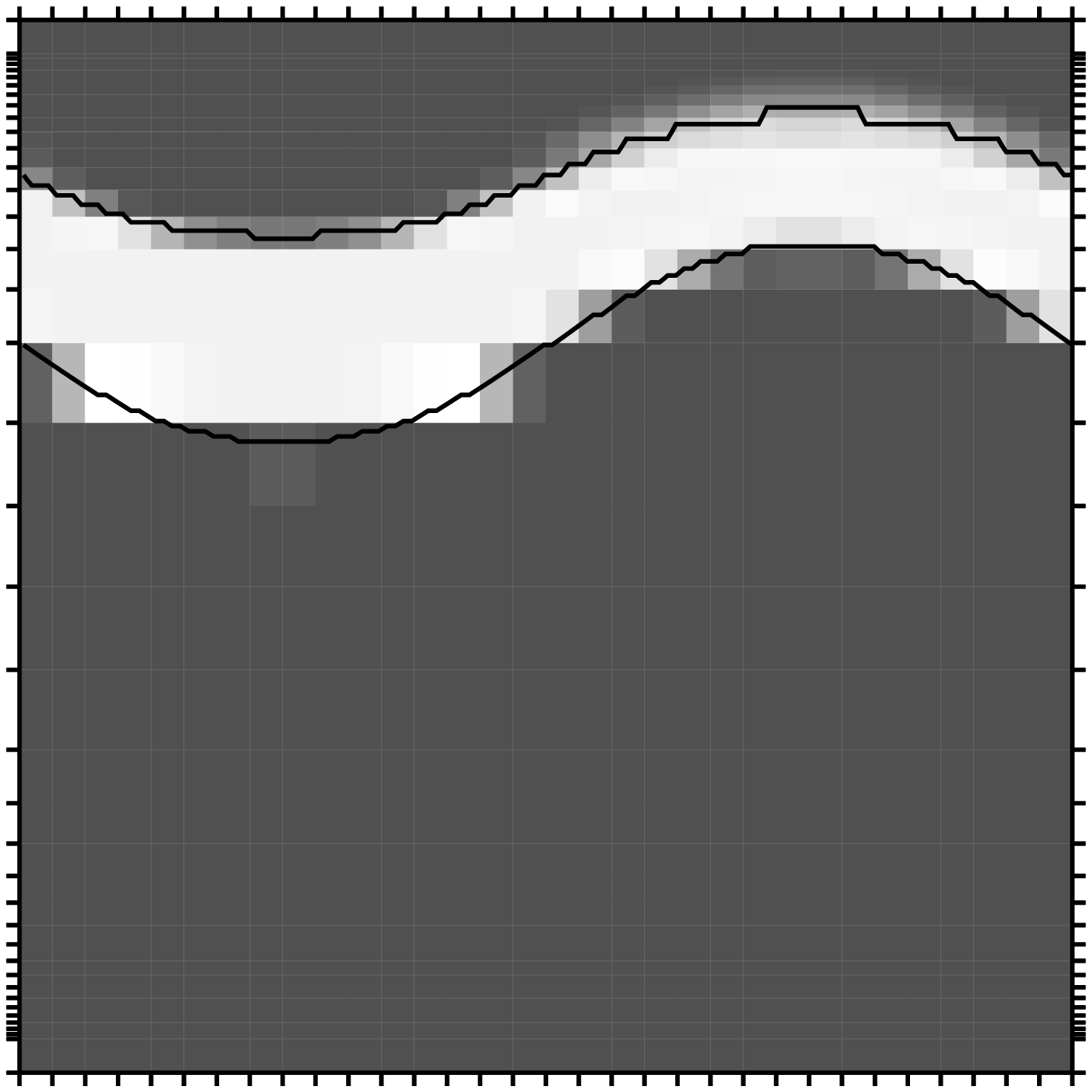}
\epsfxsize=4cm\epsfysize=4cm\epsfbox{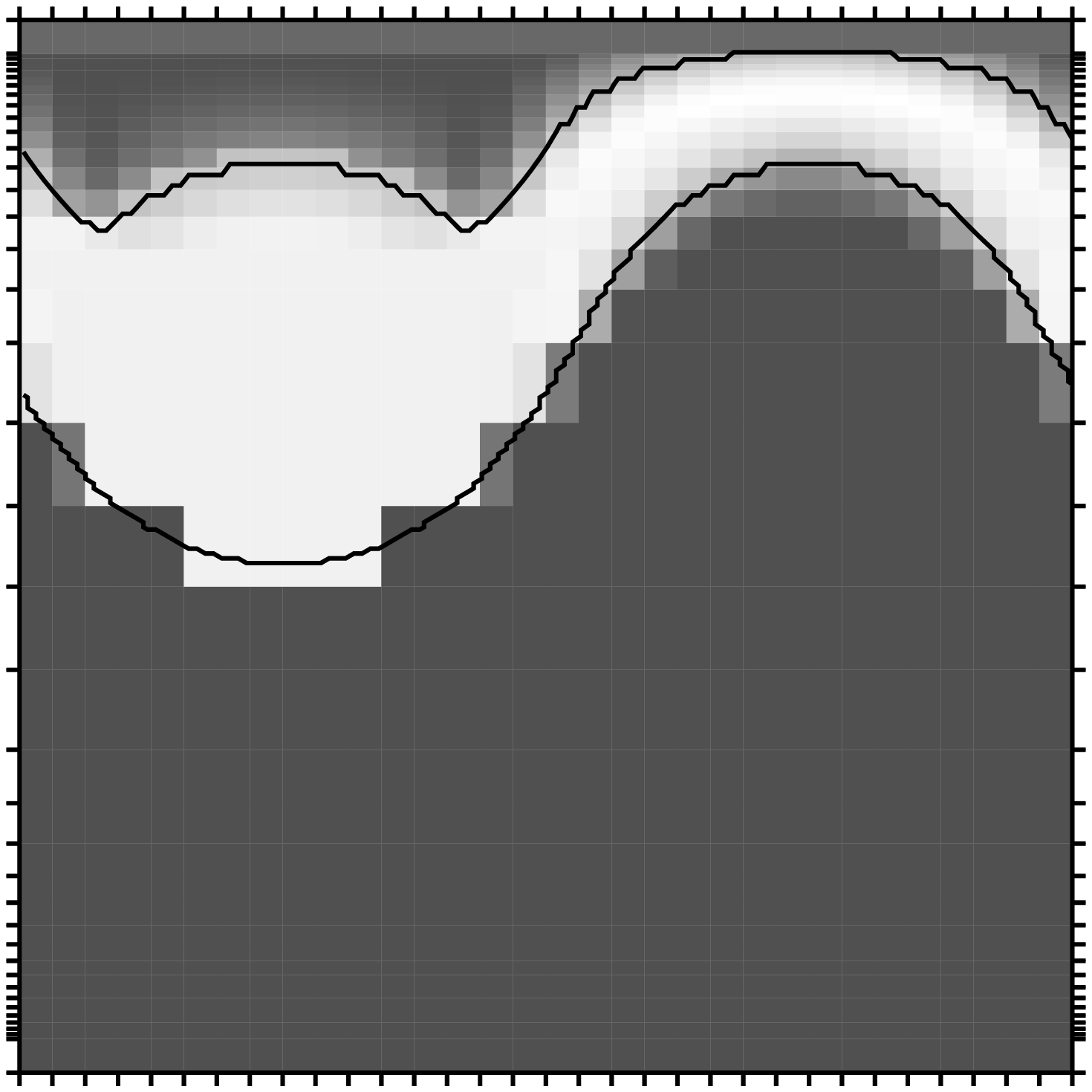}
\epsfxsize=4cm\epsfysize=4cm\epsfbox{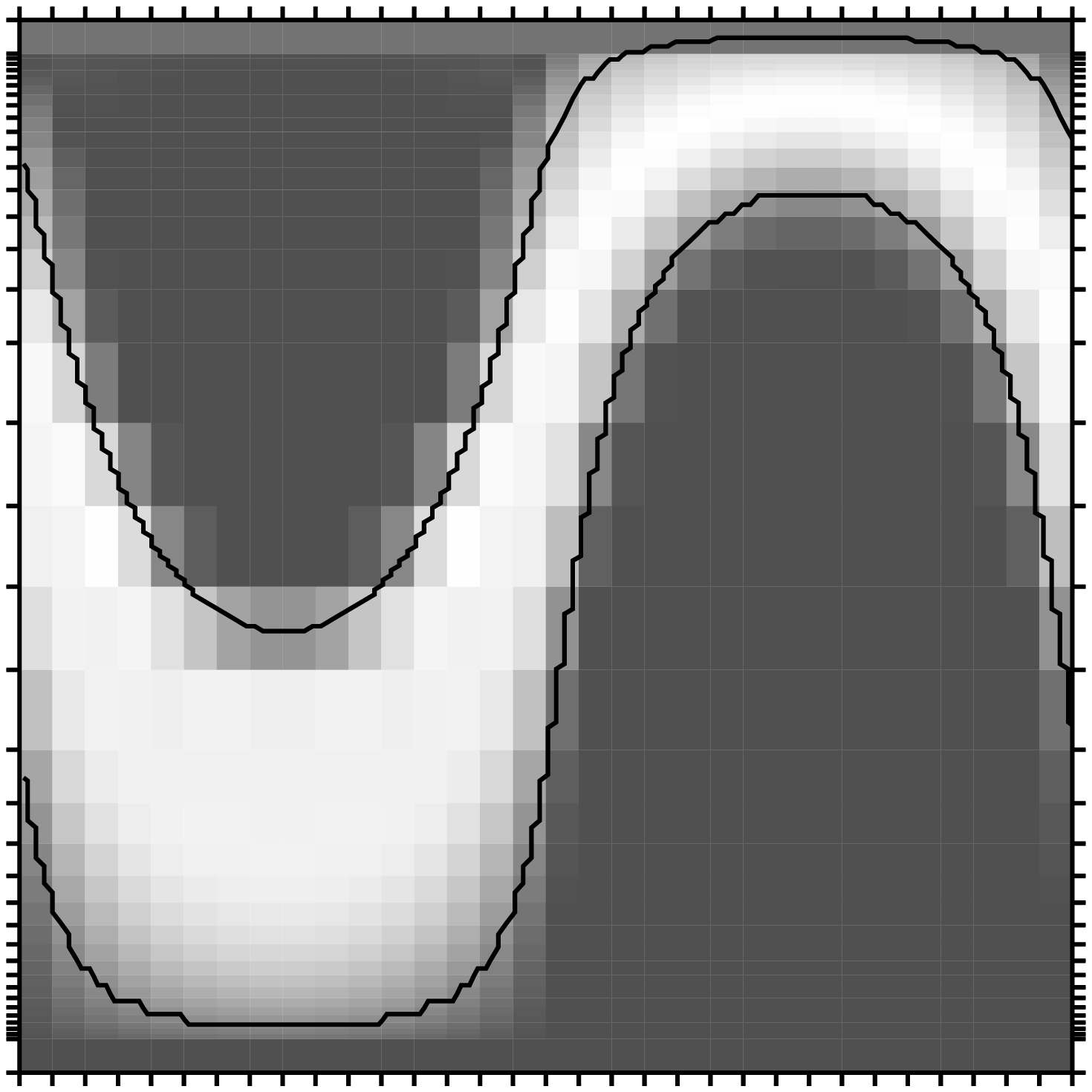}
}
\centerline{
\epsfxsize=4cm\epsfysize=4cm\epsfbox{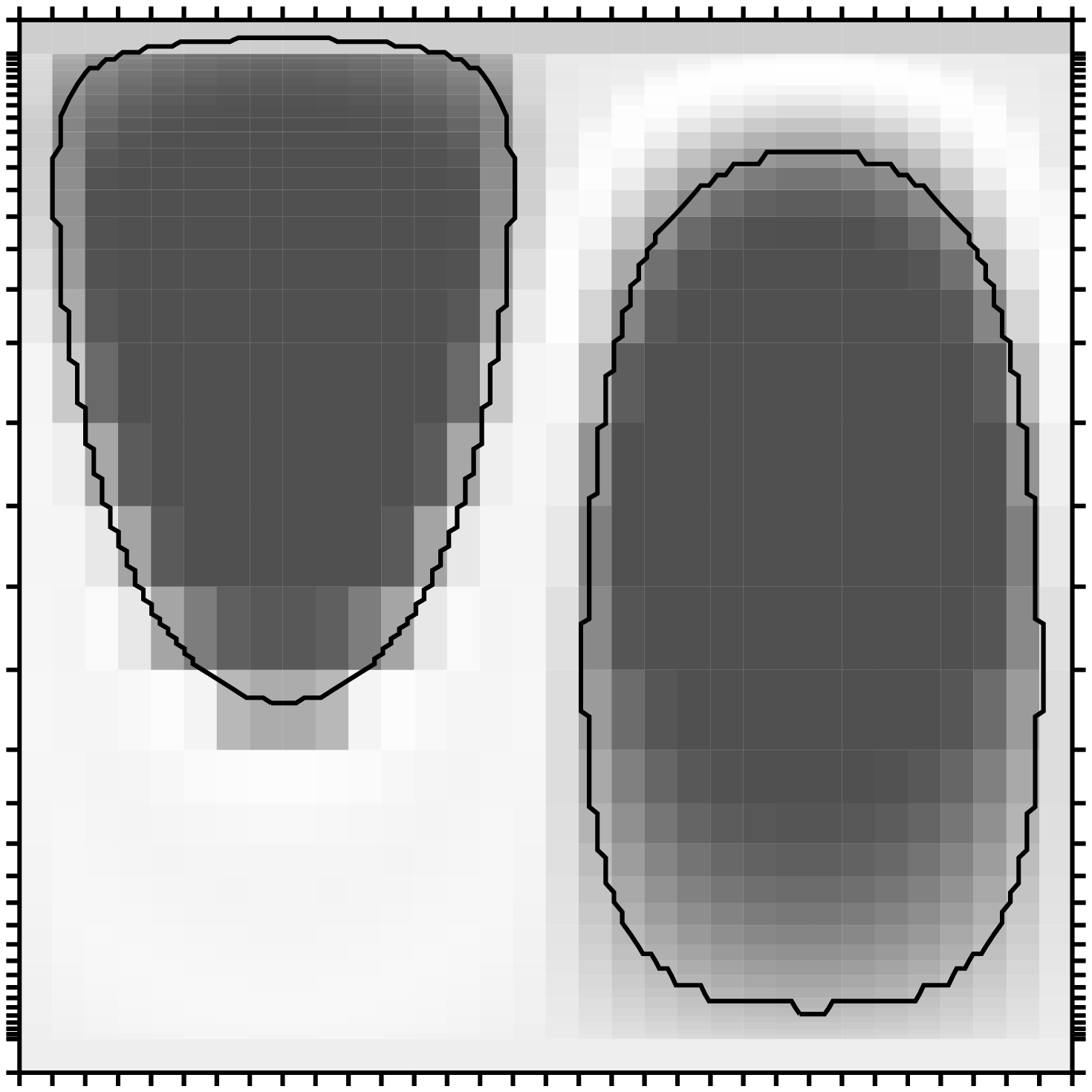}
\epsfxsize=4cm\epsfysize=4cm\epsfbox{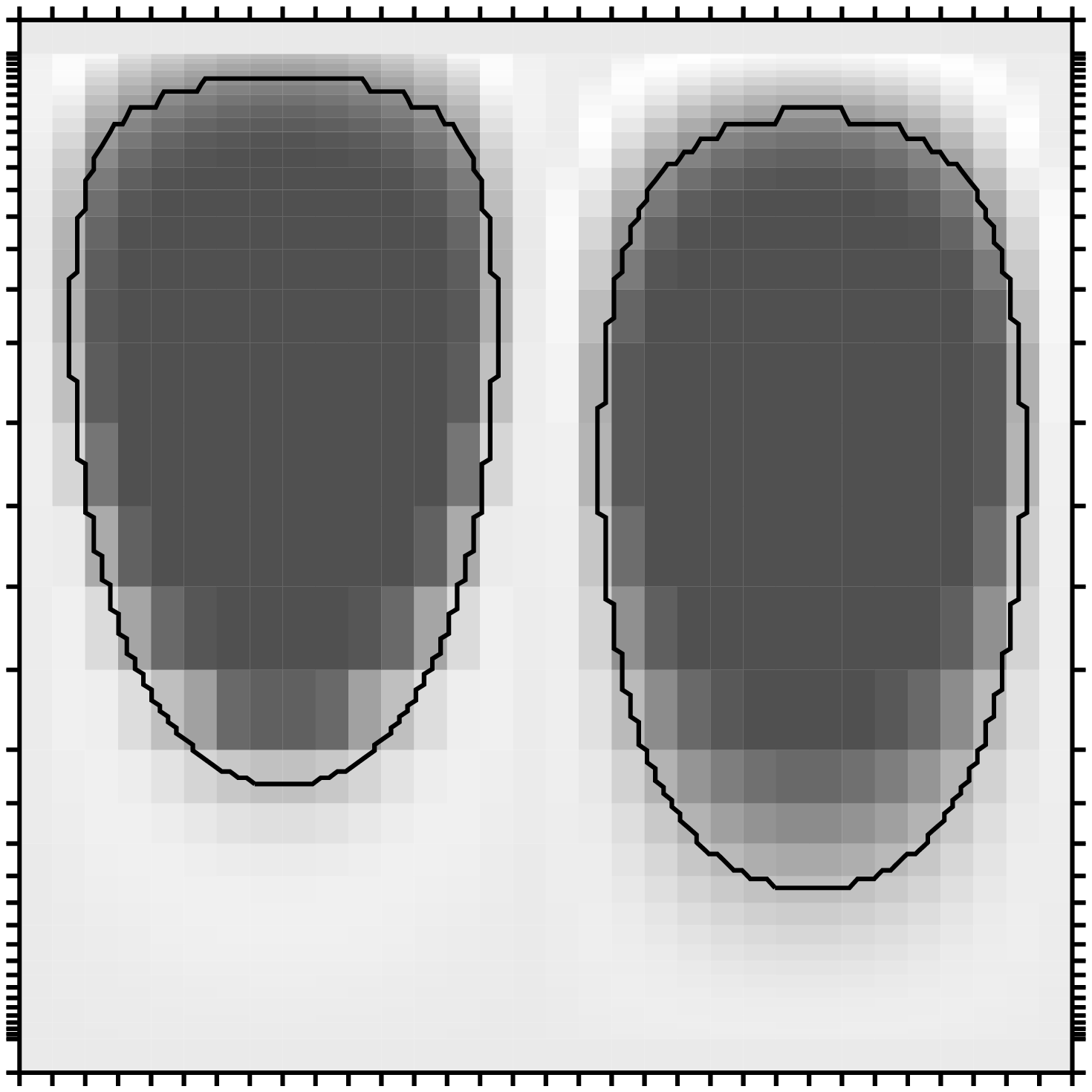}
\epsfxsize=4cm\epsfysize=4cm\epsfbox{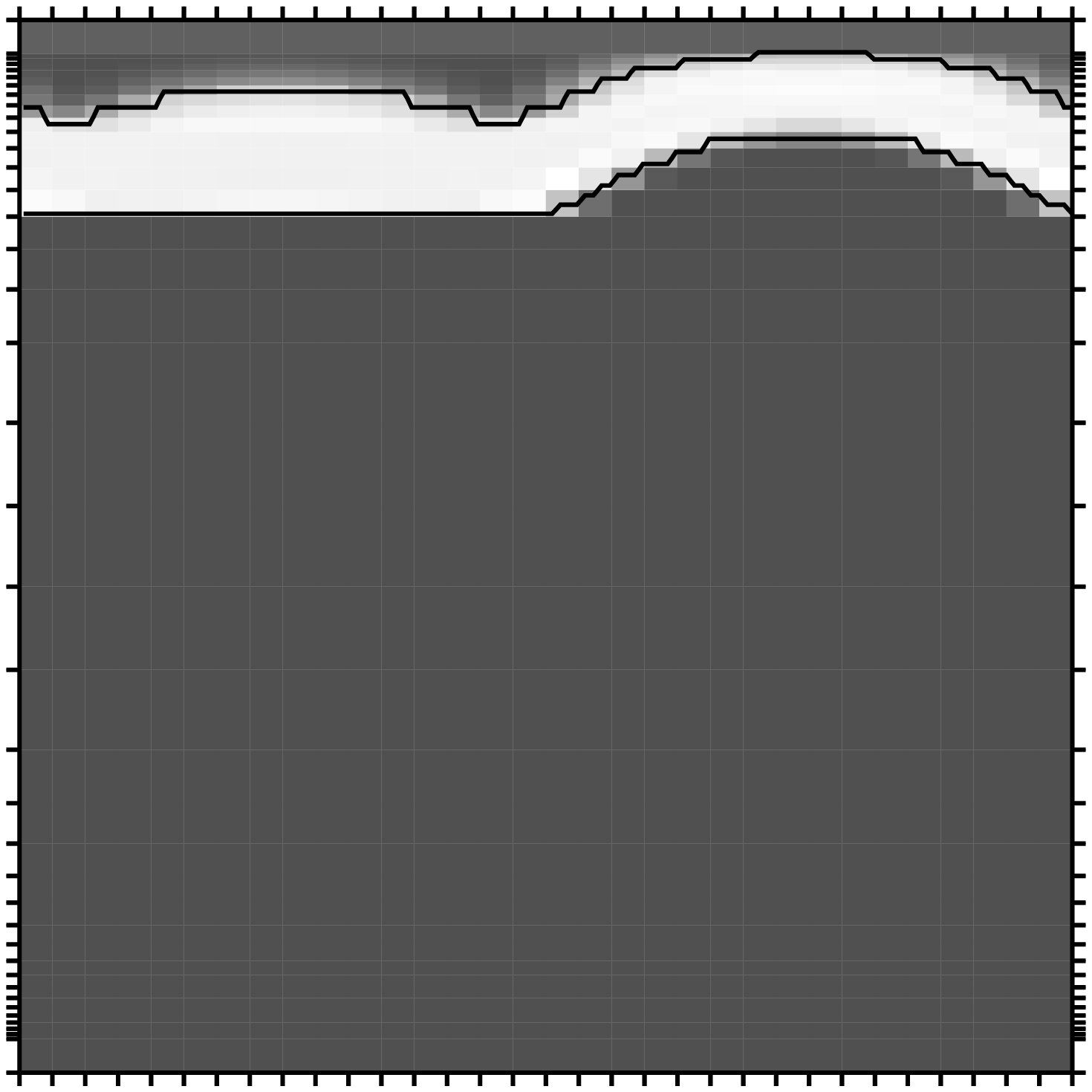}
\epsfxsize=4cm\epsfysize=4cm\epsfbox{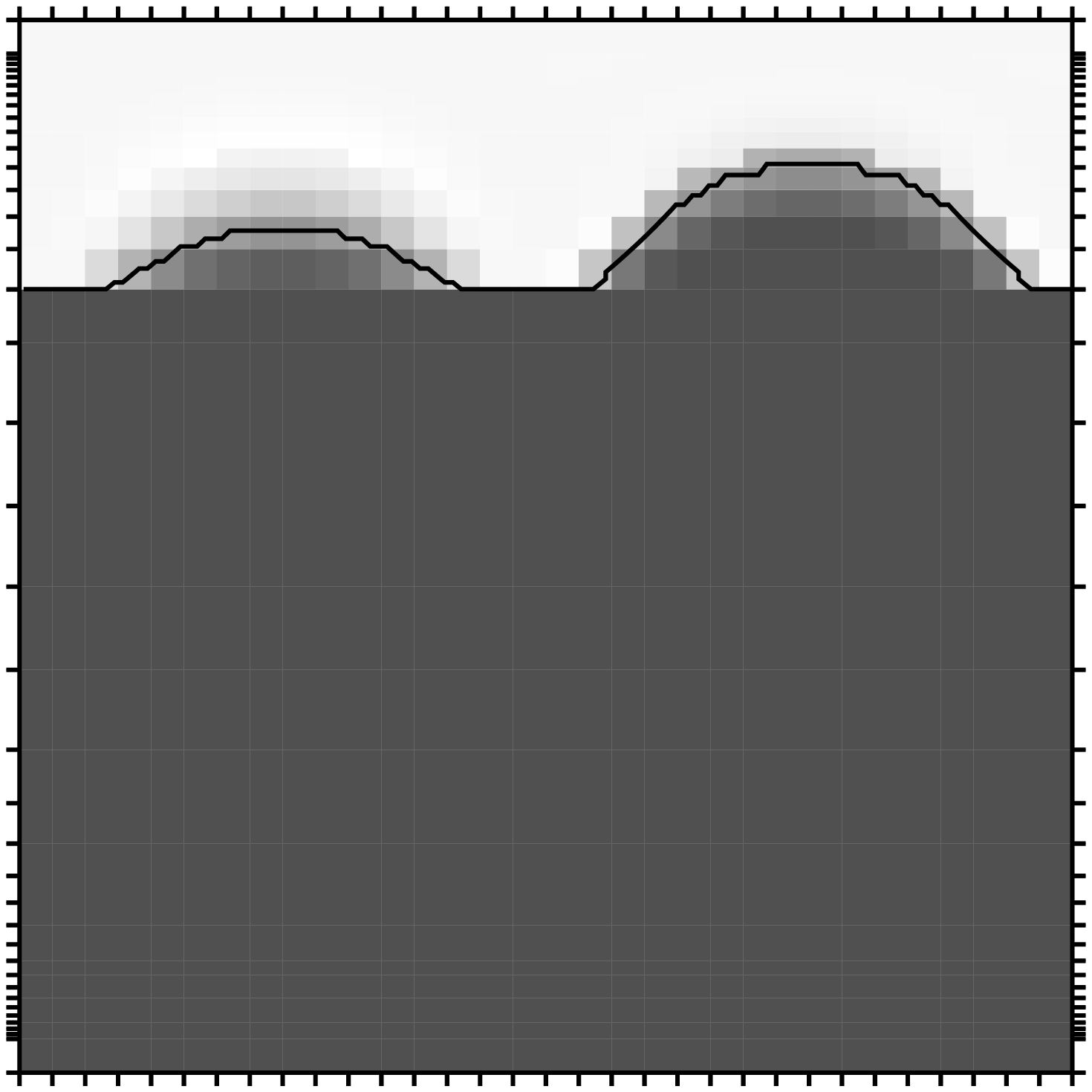}
}
\mbox{}
\caption{The radiation field $I(\mu,\phi)$ for
the test problem of the optically thick annulus at various positions in
space (from top left to bottom right at the points $A-H$ of
Fig.~\ref{fig-thick-annul-overview}).
In each panel $\phi$, from 0 to $2\pi$, is on the $x$--axis and
$\mu$, from -1 to 1, on the $y$--axis. Tickmarks on both axes are
representative of the actual values of $\mu$ and $\phi$ used in the
calculation. The image shows the
intensity resulting from the ESC transfer algorithm (the grey scale
is such that white corresponds to maximum and black to zero).
The thin solid curves mark the true contour of the object,
computed by ray--tracing with high angular resolution.
The slightly diffusive nature of the results is a consequence
of the interpolations inherent to the ESC method. Numerical diffusion
remains quite low in almost all cases, with the possible exception of
point $E$.}
\label{fig-thick-annul-radfield}
\end{figure*}}
The contours of the real images are
overplotted. The same calculation has been repeated using the LC method. The
errors of both the ESC and the LC calculations are listed in Table
\ref{tab-annul-errors}. These figures show that the errors of the ESC method
are not very much greater than those of the LC method (which result from the
discretization alone) and strenghten the reliability of the ESC algorithm.

\ifthenelse{\incltables=0}{}{
\begin{table}
\centerline{\begin{tabular}{crrrr}
Point &
$\epsilon_{ESC}$ &
$\sigma^2_{ESC}$ &
$\epsilon_{LC}$ &
$\sigma^2_{LC}$ \\
\hline
A & $ 0.060$ & $ 0.090$ & $ 0.085$ & $ 0.116$ \\
B & $ 0.034$ & $ 0.112$ & $-0.016$ & $ 0.134$ \\
C & $ 0.059$ & $ 0.112$ & $ 0.001$ & $ 0.105$ \\
D & $-0.020$ & $ 0.091$ & $ 0.002$ & $ 0.094$ \\
E & $ 0.107$ & $ 0.096$ & $-0.032$ & $ 0.084$ \\
F & $ 0.109$ & $ 0.103$ & $ 0.028$ & $ 0.089$ \\
G & $ 0.029$ & $ 0.051$ & $-0.019$ & $ 0.068$ \\
H & $-0.012$ & $ 0.056$ & $-0.046$ & $ 0.104$ \\
\end{tabular}}
\caption{The errors of ESC and the LC algorithms for the optically thick
test.
\label{tab-annul-errors}}
\end{table}
}
\subsection{Optically thin annulus}
Contrary to the optically thick case of the previous subsection,
the radiation field from an optically thin annulus cannot be
determined by a simple analytic formula such as
Eq.(\ref{eq-projection-on-sky}).  At large radii, however, the mean
intensity should follow the $1/R^2$ law and be independent of
$\Theta$. We can verify if the solution produced by the ESC algorithm
indeed has this expected
behavior. All details are the same as in the previous test, 
with the only difference that now the (constant) absorption is taken 
$10^{-6}$.

We focus on the behavior at large radii.  In radial streaming
$H(R_i,\Theta_j) \simeq J(R_i,\Theta_j)$ which, for an optically thin
source, is simply proportional to the volume integral of the emissivity
$j(R,\Theta)$ and is independent of $\Theta$
\begin{equation}
J(R) = \frac{1}{4\pi R^2}\int j(R',\Theta')(R')^2
dR'\sin\Theta'd\Theta'\, .
\end{equation}
Although close to the source the mean intensity is still very dependent on
$\Theta$, at $R\gg R_M$ the code should be able to recover the correct
behaviour $J\propto R^{-2}$ at large radii. The mean intensity
$J(R_i,\Theta_J)$ resulting from the ESC transfer calculation is shown in
Fig.~\ref{fig-jrt-surf}, where it is multiplied with $R^2$. It can be
clearly seen that the while $J$ depends on $\Theta$ for small radii, it
becomes almost independent on $\Theta$ as the radius increases and that the
inverse square law is very well reproduced.  The dependence of $J$ on
$\Theta$ at the largest radius is shown in Fig.~\ref{fig-j-at-inf}. Typical
errors are $\lesssim 5\%$, which lies within the error expected from the
coarseness of the grid.
\ifthenelse{\figs=0}{}{
\begin{figure}
\mbox{}
\centerline{\epsfxsize=8cm
\epsfysize=6.5cm
\epsfbox{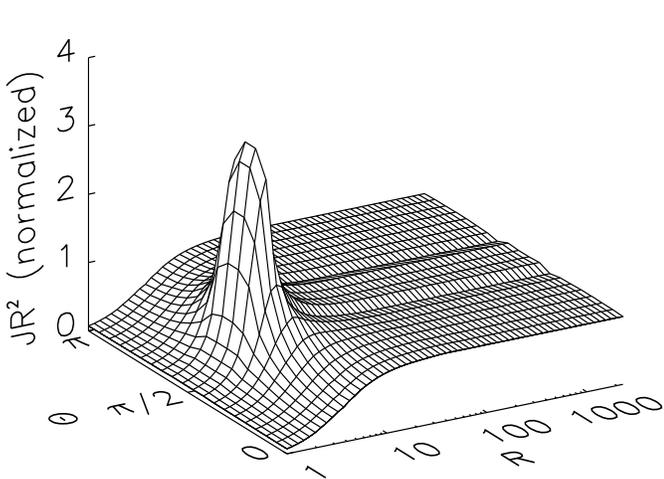}}
\mbox{}
\caption{A surface plot of the mean intensty $J(R,\Theta)$ as produced by
the ESC algorithm for the optically thin annulus. In order to show the
behavior at large radii, the mean intensity is multiplied by $R^2$ and
normalized for convenience.}
\label{fig-jrt-surf}
\end{figure}}
\ifthenelse{\figs=0}{}{
\begin{figure}
\mbox{}
\centerline{\epsfxsize=7cm
\epsfysize=5cm
\epsfbox{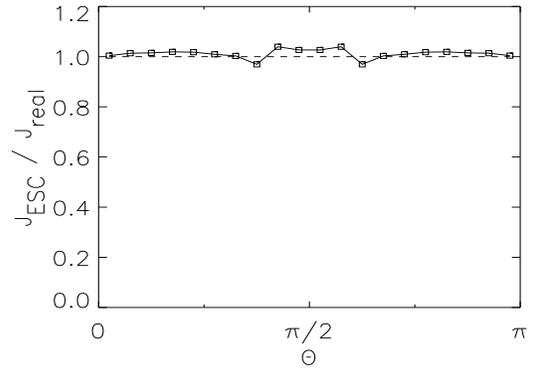}}
\caption{The computed (solid line) and the exact (dashed line) mean
intensity for the optically thin annulus, at the largest radius as a
function of $\Theta$; errors are $\lesssim 5\%$.  }
\label{fig-j-at-inf}
\end{figure}}
\ifthenelse{\figsctrans=0}{}{
\begin{figure*}
\mbox{}
\centerline{
\epsfxsize=5cm\epsfysize=5cm\epsfbox{intang_sc_trans_12_1.ps}
\epsfxsize=5cm\epsfysize=5cm\epsfbox{intang_sc_trans_12_4.ps}
\epsfxsize=5cm\epsfysize=5cm\epsfbox{intang_sc_trans_12_7.ps}
}
\centerline{
\epsfxsize=5cm\epsfysize=5cm\epsfbox{intang_sc_trans_12_10.ps}
\epsfxsize=5cm\epsfysize=5cm\epsfbox{intang_sc_trans_5_8.ps}
\epsfxsize=5cm\epsfysize=5cm\epsfbox{intang_sc_trans_5_9.ps}
}
\centerline{
\epsfxsize=5cm\epsfysize=5cm\epsfbox{intang_sc_trans_5_10.ps}
\epsfxsize=5cm\epsfysize=5cm\epsfbox{intang_sc_trans_17_6.ps}
\epsfxsize=5cm\epsfysize=5cm\epsfbox{intang_sc_trans_15_10.ps}
}
\mbox{}
\caption{Test \testannulthin{}. The radiation field at points A..I
for the optically thin annulus test case. The contours outline the
annulus.}
\label{fig-thin-annul-radfield}
\end{figure*} }
\subsection{Spherically symmetric test problem}
Although the above tests show that the ESC algorithm performs well on its
own, they are not enough to prove that it will produce accurate and reliable
results when applied to, for instance, a non--LTE line transfer
computation. Unfortunately it is not easy to test this, because to our
knowledge there exists no useful benchmark test case yet for 2--D
axisymmetric radiative transfer in circumstellar clouds.

The least we can do is to test our 2--D algorithm on a 1--D spherically
symmetric test case, and check the output against that produced by an
independent 1--D transfer calculation. This way we can at least test two of
the special features of the ESC algorithm: the additional $\mu$-points close
to $\mu\simeq 0$, and the special treatment of the intensity near $\mu\simeq
1$. These are features that are not particularly related to 2-D, and can
therefore also be tested in 1-D just as well. 

Our test cloud is a spherically symmetric power law model with hydrogen
density specified by
\begin{equation}
N_{\Hmolec}(R) = N_{\Hmolec}^0\;\left( \frac{R_0}{R} \right)^{2}\quad \cm^{-3}\,,
\end{equation}
where $R$ is the radius in $\cm$, and $N_{\Hmolec}^0=2.0\times
10^{7}\,\cm^{-3}$ is the number density at $R=R_0\equiv 1.0\times
10^{15}\,\cm$. We take a constant kinetic temperature $T_{\kin}(R) =
20\,\kel$.  The abundance of our molecule is also a constant, $X_{\molec}(R)
\equiv N_{\molec}(R)/N_{\Hmolec}(R) = 1.0\times 10^{-6}$. The systematic
velocity is taken zero. The models are computed in spherical coordinates, in
the domain $R_{\innn}\le R \le R_{\outtt}$. We take $R_{\innn}=1.0\times
10^{15}\,\cm$ and $R_{\outtt} = 7.8\times 10^{18}\,\cm$.  At the inner
boundary we put a reflective boundary condition. The incoming radiation at
the outer boundary is the $T=2.728\;\kel$ microwave background radiation.

We choose a fictive 2-level molecule which is specified by
\begin{xalignat}{1}
E_2-E_1 &= 6.0\; \cm^{-1} = 8.63244 \;\kel \\
g_2/g_1 &= 3.0 \\
A_{21}  &= 1.0\times 10^{-4}\; \sec^{-1} \\
K_{21}  &= 2.0\times 10^{-10}\; \cm^3\,\sec^{-1}
\end{xalignat}
from which the downward collision rate follows:
$C_{21}=N_{\Hmolec}K_{21}\;\sec^{-1}$.  The total (thermal+turbulent) line
width $\lwtot$ is $\lwtot = 0.150\; \km\,\sec^{-1}$ (see
Eq.~\ref{eq-turb-lineprof-def}).

The test problem presented here has high optical depth and a very
sub-critical density. It is therefore well suited to test whether 
non-LTE effects are properly computed.

The line transfer is computed in a passband of 40 frequency points equally
spaced between $-0.40\;\km\,\sec^{-1}$ and $+0.40\;\km\,\sec^{-1}$. The
$\mu$ angle is discretized using $43$ points, arranged according to the
tangent-ray prescription of Eq.(\ref{eq-mu-grid-tangent-ray}) with $3$
additional $\mu$-points around $\mu=0$ on each side. Our convergence
criterion is simply:
\begin{equation}
\max(\delta n_i/n_i) < 1\times 10^{-4}
\end{equation}
at all radii. For the radius we use an equally spaced logarithmic grid with
$(R_{i+1}-R_i)/R_i\equiv \Delta R/R=0.1720$.  We perform 4 runs: Lambda Iteration
and Accelerated Lambda Iteration with and witout Ng acceleration.

The results for the upper level population is plotted in
Fig.~\ref{fig-line-benchtest-result-low}
and compared to the results obtained independently with
SIMLINE written by V.~Ossenkopf (\cite{ossenkopfsimline:1999}).
The convergence plots for four different methods are shown in
Fig.~\ref{fig-line-benchtest-conv}.

\ifthenelse{\figs=0}{}{
\begin{figure}
\mbox{}
\centerline{
\resizebox{7.cm}{7.cm}{
\includegraphics{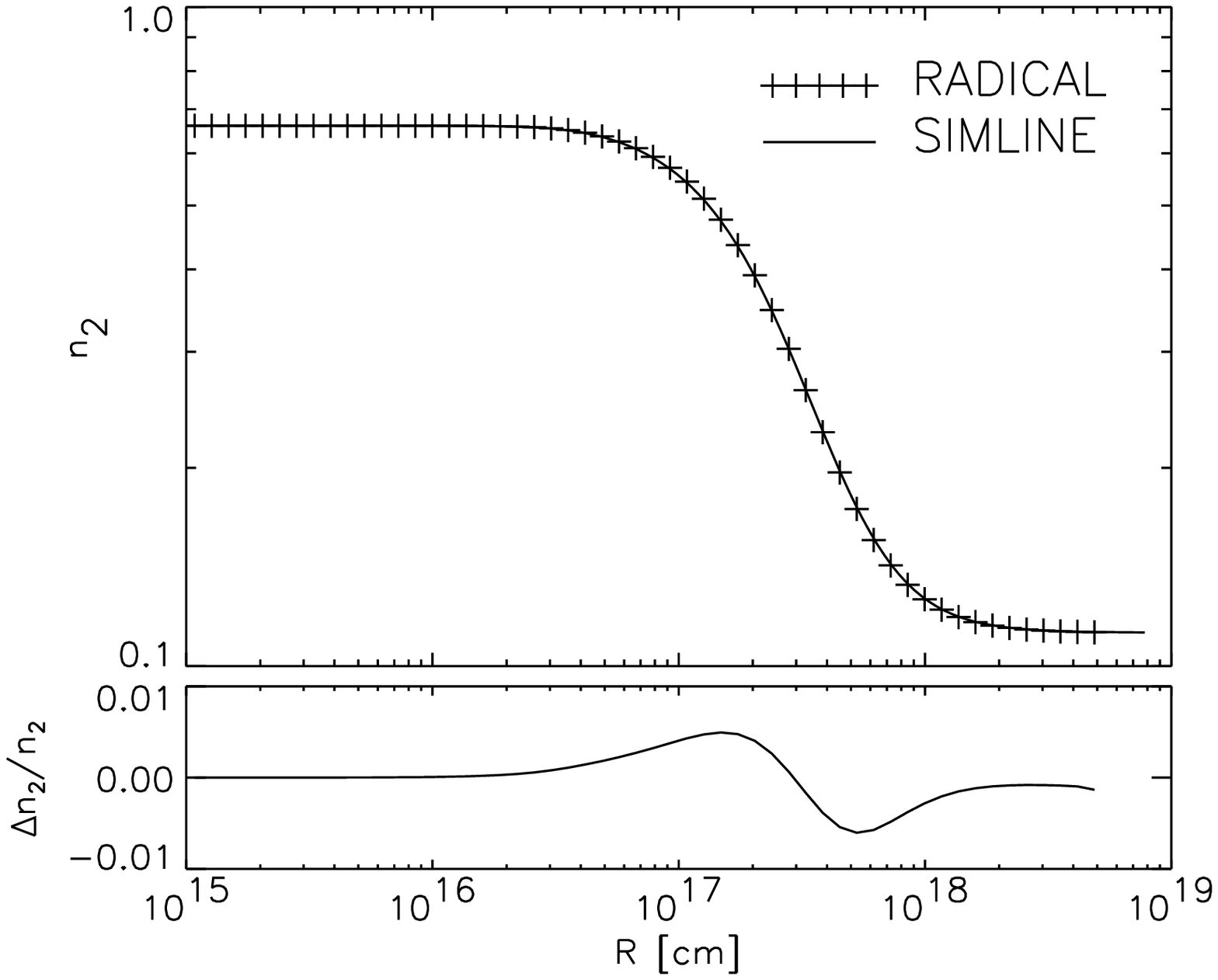}}
}
\mbox{}
\caption{The fractional population of the upper level of the 1-D test
  problem. The symbols are the solution produced by RADICAL, which is our
  code based on the ESC algorithm. ALI+Ng were used. The solid line is the
  solution found by the program SIMLINE, which is an independent 1-D line
  transfer program written by V.~Ossenkopf (\cite{ossenkopfsimline:1999}). The
  difference between the two solutions (normalized to the SIMLINE solution)
  is shown in the lower panels.}
\label{fig-line-benchtest-result-low}
\end{figure}}
\ifthenelse{\figs=0}{}{
\begin{figure}
\mbox{}
\centerline{
\resizebox{7.5cm}{6.5cm}{
\includegraphics{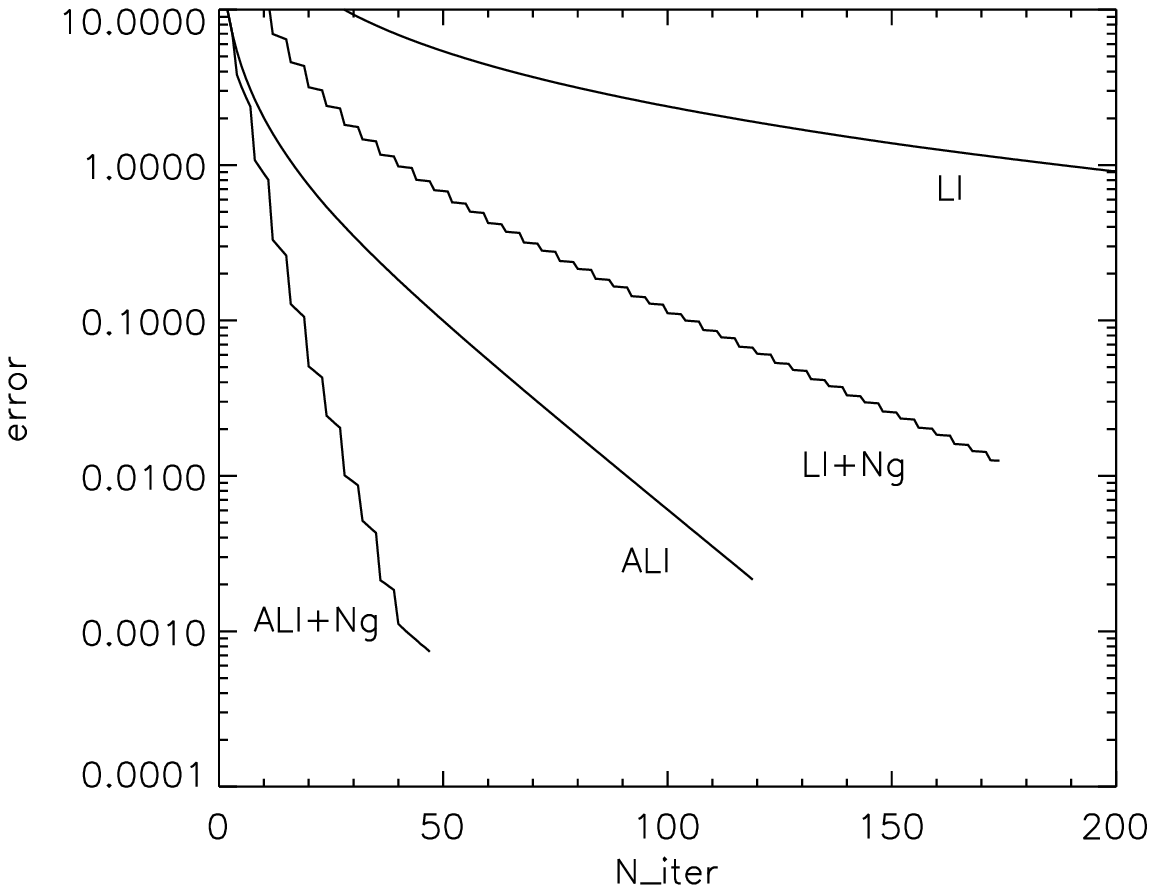}}
}
\caption{Convergence plot for the 1-D test problem. The largest error
is plotted against the iteration number for four different methods. We
define the error as the maximum relative difference between the ``real''
solution and the level populations at iteration $N_{\iter}$. The ``real''
solution was obtained with ALI+Ng and converged down to $\max(\delta
n_i/n_i)<10^{-6}$, which is a 100 times stricter convergence criterion than
the last iteration step plotted for ALI+Ng in this figure.}
\label{fig-line-benchtest-conv}
\end{figure}}

More test problems, and a more extensive discussion of them was presented
by Dullemond (\cite{dullemondthesis:1999}).

\section{A simple model for the Egg Nebula}\label{sec-eggneb}
Now that the algorithm has been tested, we demonstrate here how it can be
used in practice. Our first example is a simple model of the optical
appearance of the Cygnus Egg Nebula (CRL 2688). This object has been
extensively studied ever since its discovery by Ney et
al.~(\cite{ney:1975}). It is a bipolar reflection nebula surrounding an F5
supergiant of $T_{\eff}=6500 \kel$ (Crampton \cite{crampton:1975}). At
optical wavelength it appears as a diffuse bi--lobed nebula with two sharply
edged ``searchlight beams'' emerging from each of the poles (Sahai et
al.~\cite{sahaitraug:1998}).  The lobes are separated by a dark equatorial
lane which completely obscures the central star.

The optical emission from this nebula can be understood as reflected
starlight escaping from the nebula through polar cavities (Latter et
al.\cite{latterhora:1993}, Morris \cite{morris:1981}). It is clear that the
``searchlight beams'' are due to single scattering of direct starlight by
dust grains. The lobes are, however, more likely to be the result of
multiple--scattering.

We will model this multiple--scattering process in 2--D with the MESC
algorithm, in an attempt to reproduce the complex optical appearance
of the nebula. Our setup consists of an almost spherical wind with a
cavity at both poles. The density in the cavities is small, but it is
still high enough to reflect sufficient amounts of starlight. To
reproduce the twin--beams at both poles, we place a small blob of
matter at the polar axis in the cavity, causing a shadow.  A star is
placed at the center of the coordinate system to illuminate the nebula
from within.

We model only a single frequency in the optical, at 600 nm. For this reason
we refrain from taking actual realistic dust opacities, and specify total
optical depth and scattering albedo instead. The dust density is shown as
contour plots in Fig.~\ref{fig-eggdens}. The total optical depth at the
equator is about 60. The ratio of absorption over scattering is $0.3$. The
dust scattering is assumed to be isotropic, which suffices for the present
simplified example. We do not specify the dust temperature for this setup
since thermal emission at 600 nm is negligible.

The simulation was performed by RADICAL, using the MESC algorithm, and
applying Accelerated Lambda Iteration and Ng acceleration. The image
subsequently produced by formal integration is shown in
Fig.~\ref{fig-eggneb-revgrey}. The model reproduces the searchlight beams
and the diffuse glow. It also naturally reproduces the intensity difference
between the north and south lobe. This is a result of the slight inclination
at which the object is seen. For light emerging from the south lobe, the
path length through the outer regions of the nebula is larger than for the
north lobe.  \ifthenelse{\figs=0}{}{
\begin{figure}
\mbox{}
\centerline{\epsfxsize=7cm
\epsfysize=7cm
\epsfbox{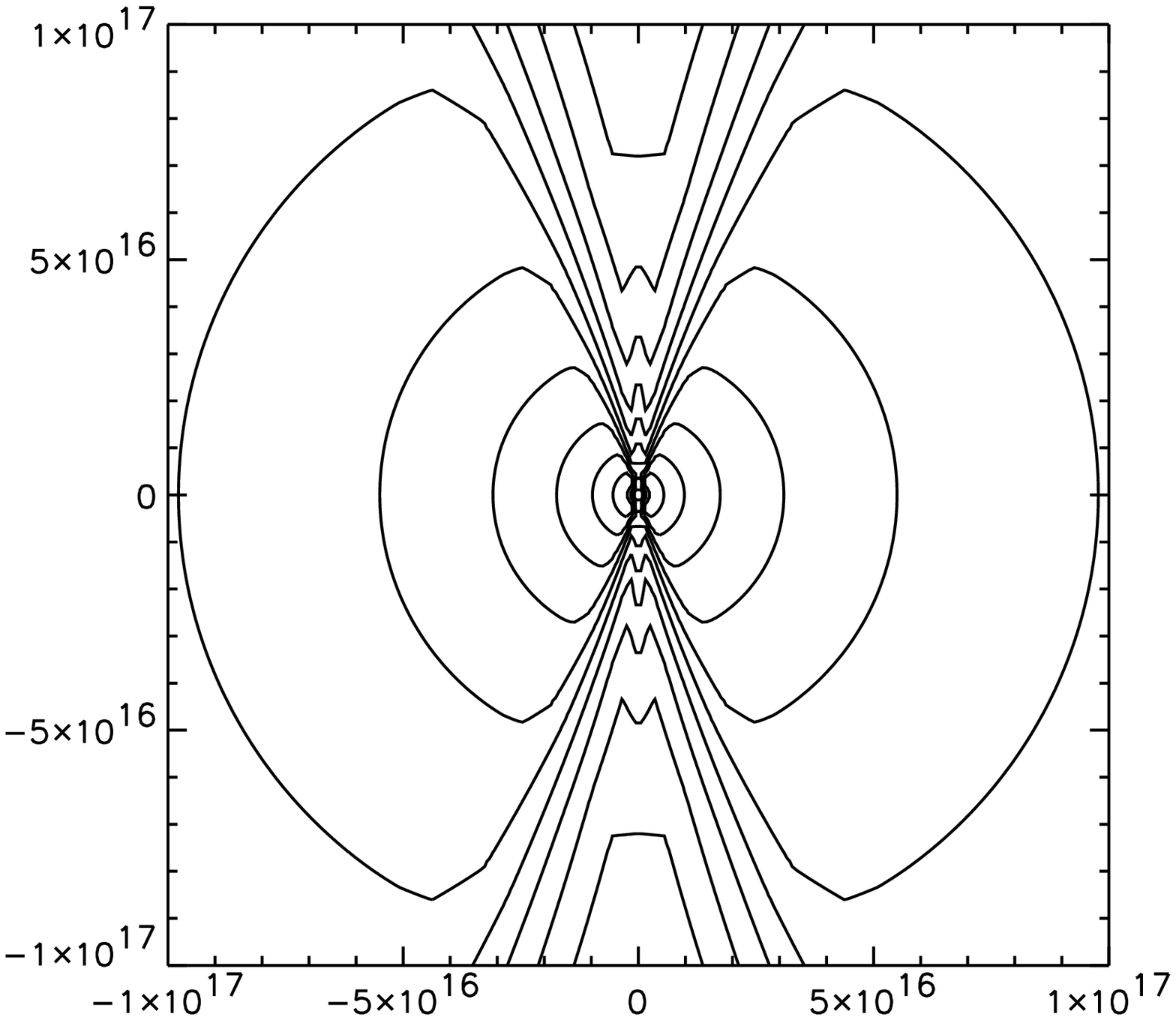}}
\mbox{}
\caption{The density contours for the dust envelope around the central
star of the Egg Nebula. Contours are logarithmically spaced, separated
by factors of $\sqrt{10}$. The distance scale on the x- and y-axis is
in centimeters. At the polar axis one can see a small blob of matter.
This blob is responsible for the shadow.}
\label{fig-eggdens}
\end{figure}}
\ifthenelse{\figs=0}{}{
\ifthenelse{\bigegg=0}{
\begin{figure}
\mbox{}
\centerline{\epsfxsize=8cm
\epsfysize=8cm
\epsfbox{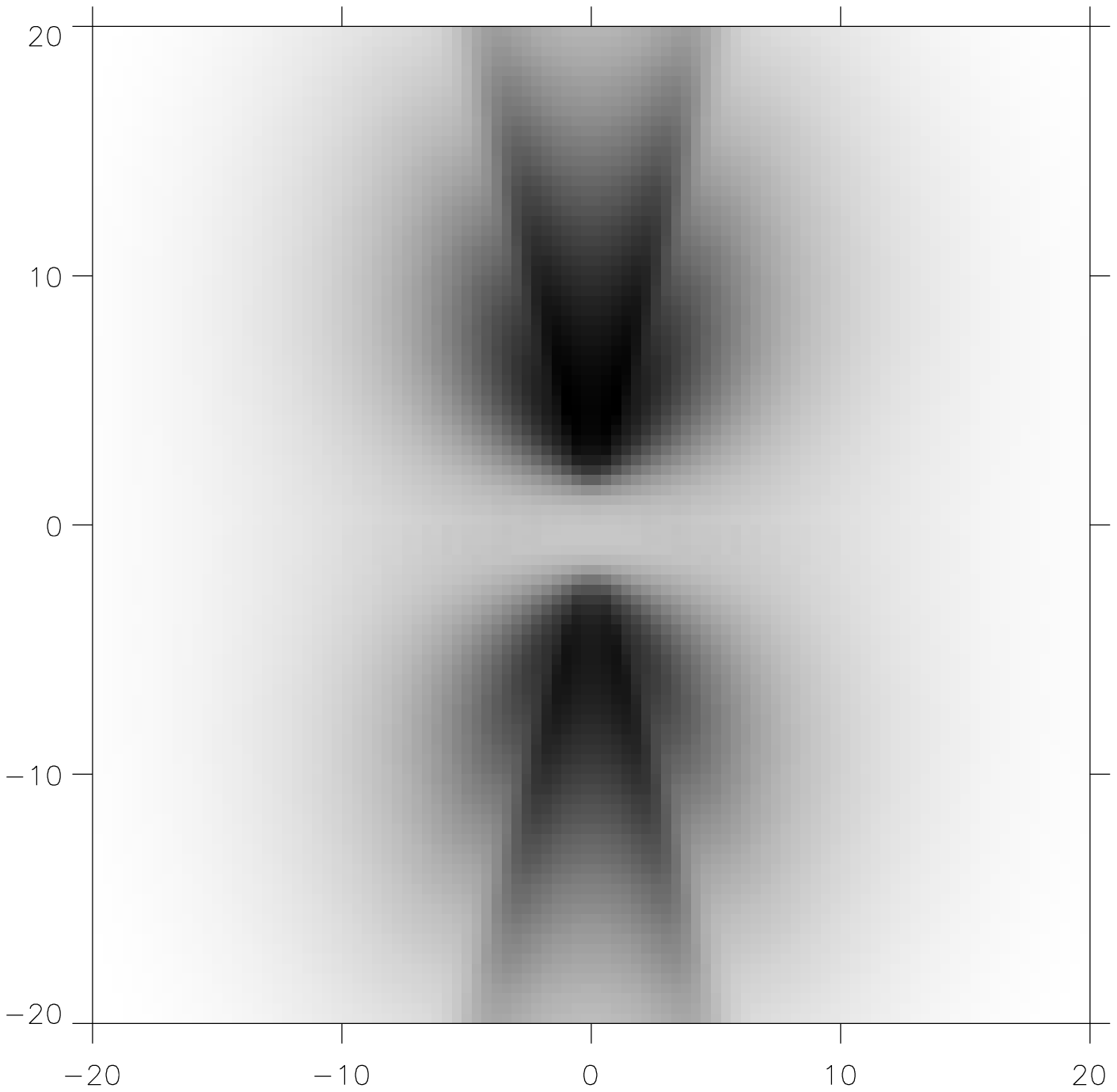}}
\mbox{}
\caption{The synthetic image of the Egg Nebula, produced by RADICAL (our
code based on the ESC algorithm).  The horizontal and vertical scale is in
arcseconds.  The color coding is reverse logarithmic grey scale, slightly
modified (up to 15\%) to enhance contrast in the lobes. }
\label{fig-eggneb-revgrey}
\end{figure}}
{
\begin{figure*}
\centerline{\epsfxsize=14cm
\epsfysize=14cm
\epsfbox{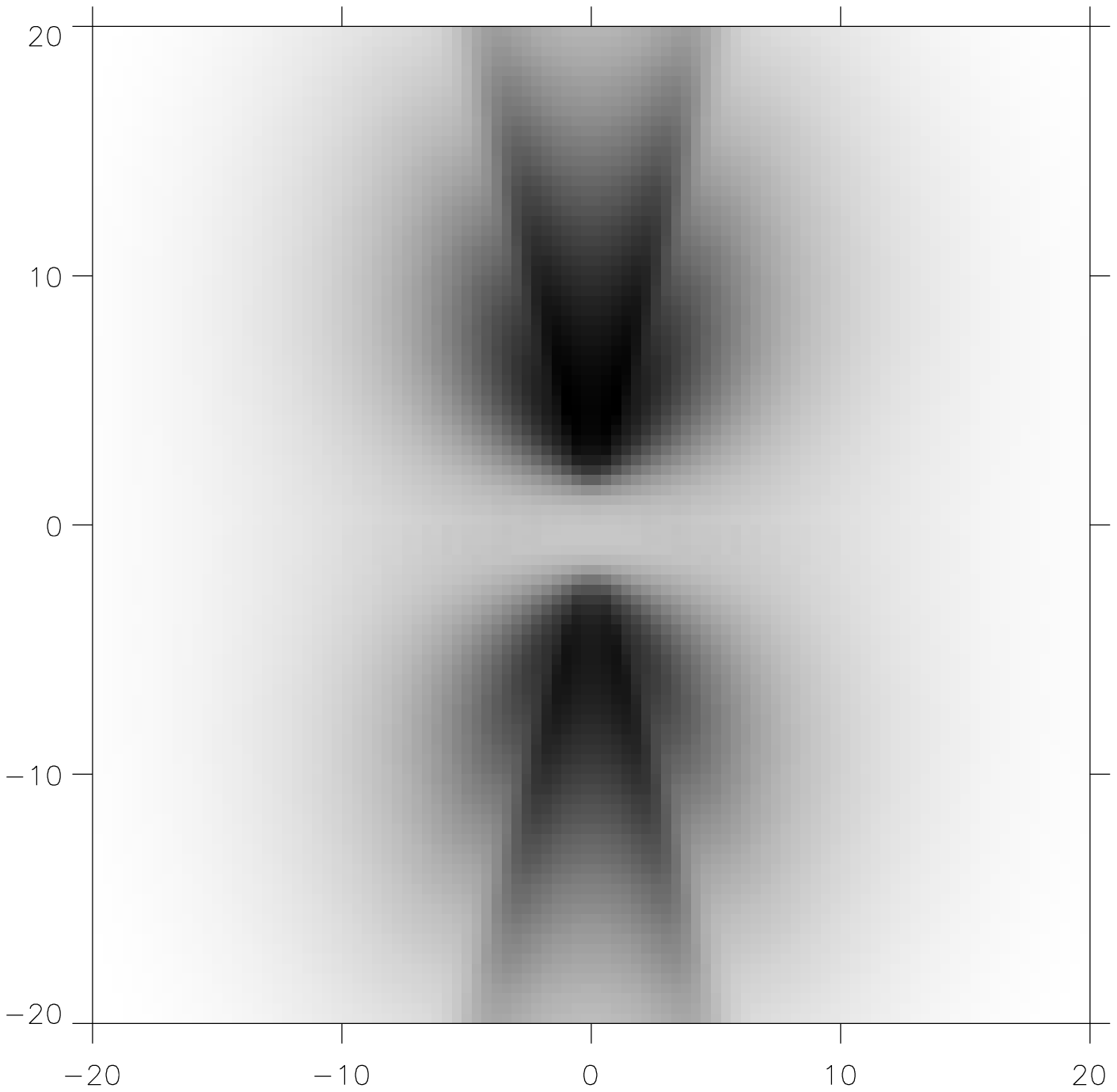}}
\caption{The synthetic image of the Egg Nebula, produced by RADICAL.
The color coding is reverse logarithmic grey scale. The color table
was slightly modified to enhance contrast in the lobes, but the
modifications remain within 15\%.}
\label{fig-eggneb-revgrey}
\end{figure*}}
}

Although the model resemblances the HST image of Sahai et al., it should be
noted that the density structure that we have used may not be consistent
with observations at other wavelengths.  For example, HCN observations by
Bieging \& Nguyen-Q-Rieu (\cite{bieging:1996}) seem to rule out the presence
of a cavity in the wind. Also the rather high albedo may be difficult to
reconcile with the fact CRL 2688 is carbon--rich.

\section{A model of line transfer in a collapsing protostellar cloud}\label{sec-hcb}
In this section it will be demonstrated how the 2-D transfer
algorithm can be used in the observational study of low-mass star
formation in dense molecular cloud cores.

Low mass star formation takes place in dense molecular cloud cores.
According to the spherically symmetric model of Shu (\cite{shu:1977}), such
a core develops a cusp with density close to a $1/R^2$ powerlaw.  Once a
gravitational instability is trigged, the centeral part of the cloud
collapses, and forms a star. An expansion wave propagates into the cloud
towards larger radii, allowing more and more matter to fall supersonically
down the potential well, and add to the protostar's mass.

Both observational evidence and theoretical arguments, however, indicate
that purely spherical collapse is rare. Any slight amount of angular
momentum in the primordial cloud will cause deviations from sphericity as
centrifugal forces tend to dominate over radial infall deep down the
potential well. And even before the collapse stage these primordial clouds
often appear to be non-spherical (Myers et al.~\cite{meyersfuller:1991}).
Theoretical models of non-spherical protostellar collapse include, among
others, Ulrich (\cite{ulrich:1976}), Cassen \& Moosman
(\cite{cassenmoosman:1981}), Terebey et al.~(\cite{terebyshucas:1984}, Galli
\& Shu \cite{gallishu:1993a}.

The models of Cassen \& Moosman and Ulrich (hereafter CMU) focus on the
inner free-falling part of the collapsing cloud, and assume that the
material originates from an originally spherical cloud with some angular
momentum. Their model is almost spherical at large radii, but flattens off
closer towards the center, and forms a disk near the centrifugal radius.
This model was later extended by Hartmann et al.~(\cite{hartcalvboss:1996},
hereafter HCB) to include flattening of the parent cloud. These models show
that the inner free-fall part of an initially flattened cloud naturally
tends to form a bipolar cavity, which is often observed in YSO. These models
are distinctly non-spherical at all radii, despite the fact that centrifugal
forces only dominate at small radii. It is therefore evident that fitting
such models to molecular line observations requires 2-D axi-symmetric 
radiative transfer computations.

In this section we perform such a calculation, using the algorithms of this
paper. We solve the non-LTE level populations for the first 7 rotational
levels of HCO$^{+}$, and compute the predicted single--dish spectra.

\subsection{Description of the model}
In our HCB models we assume that the radius of the expansion wave
$\Raccr$ is outside our domain, so we shall confine our study to
the free-fall inner region of the collapsing sheet-like molecular
cloud. We assume that matter in the parent cloud had a small amount
of rotation in the plane of the sheet before it collapsed. According
to the HCB model, the velocity field of the gas is given
by the formulae of Ulrich (\cite{ulrich:1976}):
\begin{xalignat}{1}
v_R &= - \left( \frac{GM}{R} \right)^{1/2} \left( 1 + \frac{\mu}{\mu_0} \right)^{1/2}\\
v_{\Theta} &= \left( \frac{GM}{R} \right)^{1/2} \left( \frac{\mu_0-\mu}{\sqrt{1-\mu^2}} \right)
\left( 1 + \frac{\mu}{\mu_0} \right)^{1/2} \\
v_{\phi} &= \left( \frac{GM}{R} \right)^{1/2} \left( \frac{1-\mu_0^2}{1-\mu^2} \right)^{1/2}
\left( 1 - \frac{\mu}{\mu_0} \right)^{1/2}
\comma
\end{xalignat}
where $\mu\equiv\cos(\Theta)$ and $\mu_0\equiv\cos(\Theta_0)$. The
angle $\Theta_0$ is the $\Theta$-coordinate that the gas parcel had
when it started its free-fall at large radius. For a given
$(R,\Theta)$ the value of $\Theta_0$ can be found from 
\begin{equation}
\frac{R}{\Rcentr}\left( 1-\frac{\mu}{\mu_0} \right) = (1-\mu_0^2) 
\comma
\end{equation}
where $\Rcentr$ is the centrifugal radius, i.e.~the radius at which centrifugal
forces equal gravity. This is the outer radius of the disk that is formed as
a result of the rotation.

The density of the gas for the HCB model is given by
\begin{equation}
\begin{split}
\rho_{HCB}(R,\Theta) =& \frac{\dot M}{4\pi(GMR^3)^{1/2}} 
\frac{\eta\sech^2(\eta\mu_0)}{\tanh(\eta)} \\
& \times\left( 1+\frac{\mu}{\mu_0} \right)^{-1/2} \left( \frac{\mu}{\mu_0}
+\frac{2\mu_0^2\Rcentr}{R} \right)^{-1} 
\end{split}
\end{equation}
where $\eta$ is a dimensionless flattening parameter, which is
roughly equal to the ratio of the accretion radius $\Raccr$ to the
sheet thickness $H$.  HCB argue that this value must be somewhere in
between $\eta=0$ and $\eta=\eta_{\maxx}\simeq 4$. For $\eta=0$ the CMU models
are reproduced. Density contours of this free-falling envelope, for
different flattening parameters, are shown in Fig.~\ref{fig-hcb-dens-eta-all}.
\ifthenelse{\figs=0}{}{
\begin{figure*}
\mbox{}
\centerline{\resizebox{13.3cm}{3.32cm}{
\includegraphics{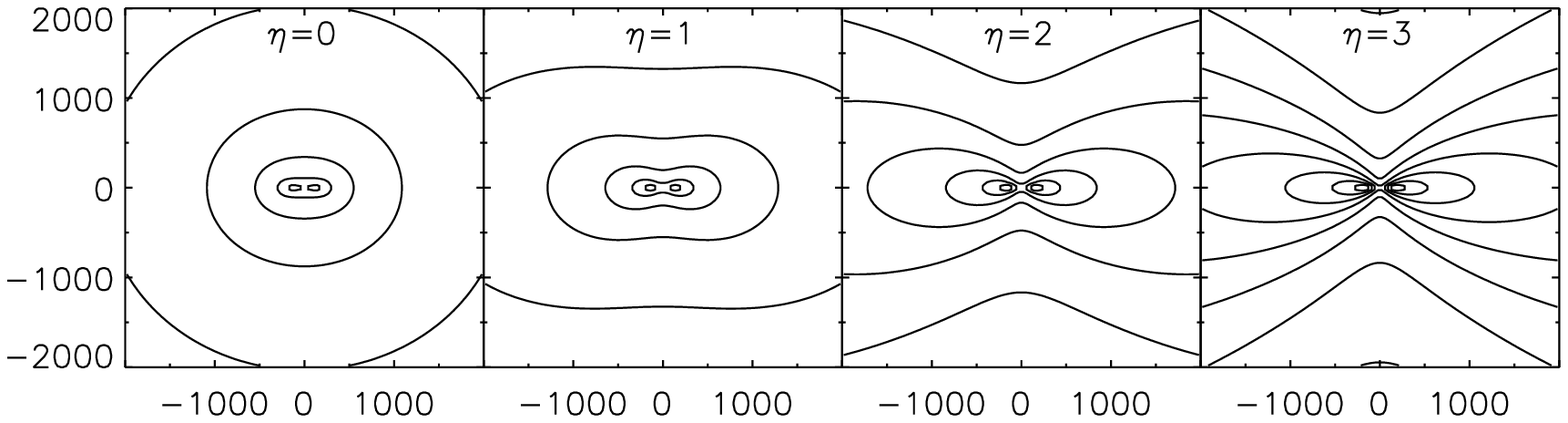}}
}
\vspace{1em}
\mbox{}
\caption{The density structure of the flattened collapsing cloud model
of HCB, for varying flattening parameter $\eta$. Scales are in $AU$. The centrifugal
radius is at $\Rcentr=100\;\AU$. The effect of rotation can be seen most
clearly in the $\eta=0$ case, which is spherical at large radius, but
flattens off near $R=\Rcentr$. 
}
\label{fig-hcb-dens-eta-all}
\end{figure*}
}
The centrifugal radius of the infalling envelope is at $\Rcentr=100\;\AU$.
We place a thin disk with a radius of $R_{\hbox{\scriptsize disk}} =\Rcentr
=100\;\AU$ at the equator. A zoom-in of the density distribution, down to
the scale of the disk, is shown in Fig.~\ref{fig-hcb-dens-zoomin}.  We will
ignore any emission from the disk, and merely treat it as a light-blocking
boundary condition at the equator.
\ifthenelse{\figs=0}{}{
\begin{figure}
\mbox{}
\hbox{\hspace{2em}
\resizebox{8cm}{4.0cm}{
\includegraphics{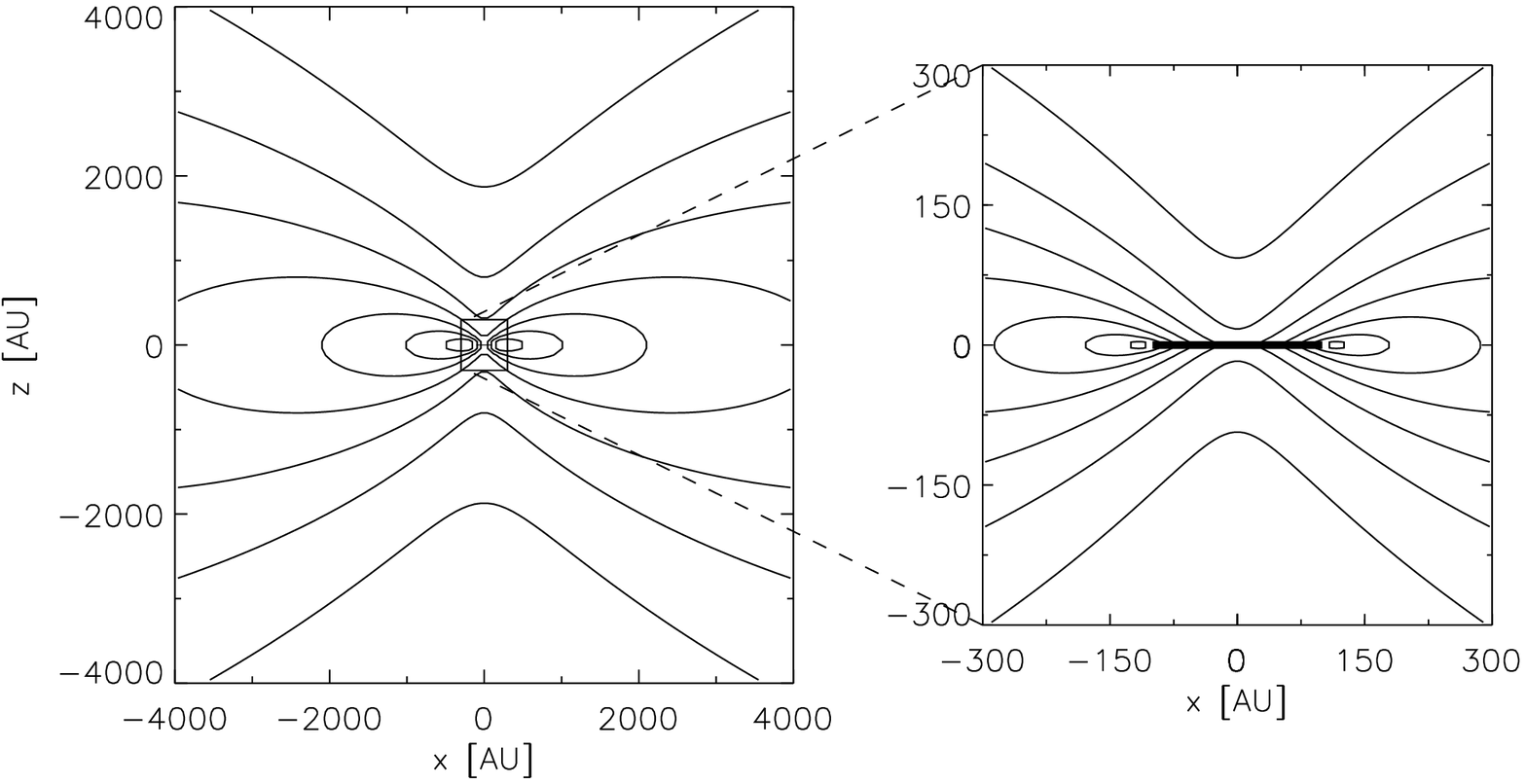}}}
\mbox{}
\caption{A zoom-in of the density structure of the $\eta=3$ case of
the HCB model. The figure on the right is a
zoom-in to the scale of the centrifugal radius. The bar represents
the disk.}
\label{fig-hcb-dens-zoomin}
\end{figure}
}
\subsection{Non-LTE line transfer}
We compute the line transfer problem for four HCB models, with
flattening parameter $\eta=0,1,2,3$ for models 1,2,3,4 respectively.
The adopted valued for the accretion rate and the turbulent line width
are $\dot M=2.4\times 10^{-6}\,M_{\odot}\yr^{-1}$ and
$\turblw=0.25\,\km\sec^{-1}$.

For these models we compute the non-LTE line transfer problem of the
lowest few rotational levels of HCO$^{+}$, including the effects of
the moving medium. We assume a cosmic background (CMB) continuum as incident radiation at
the outer edge of the computational domain. Dust emission and opacity
are neglected in the line transfer, which is justified for the
lower-lying HCO$^{+}$ lines because the nebula is optically thin to
dust in the millimeter and sub-millimeter, and radiative pumping by
dust continuum is not important for HCO$^{+}$. Also, we need not
include the dust emissivity in the computation of the emerging
spectra, since we shall show only the spectra with the dust- and
CMB-continuum removed. The radiative transfer is computed within a
range of $R\in [70,10^4]\,\AU$. As an inner
boundary we have a vacuum. The cross sections for H$_2$-HCO$^{+}$
collisional transitions were taken from Monteiro
(\cite{monteiro:1985}) and Green (\cite{green:1975}).
We adopt an HCO$^{+}$ abundance of $2\times 10^{-9}$
The gas temperature is taken to be $T=20\kel$ throughout the
cloud.

We perform the non-LTE line transfer for all four models. The resulting
non-LTE level populations for model 3 are shown in
Fig.~\ref{fig-hcb-levpop}, and the corresponding excitation temperatures
are shown in Fig.~\ref{fig-hcb-exctemp}. One can see that at the
equator ($\Theta=\pi/2$) the levels are almost thermalized, except at large
radii. This is due to the much larger density at the equator than at the
pole. The drop in excitation temperature at large radii is a result of the
decoupling of radiation and matter. At large radii the level populations
will be strongly influenced by the cosmic background radiation. Another
interesting phenomenon occurs at small radii near the pole: the excitation
temperature exceeds the gas temperature. This effect was discussed by Leung
\& Liszt (\cite{leunglist:1976}) for the CO 1-0 transition. It can be
understood as resulting from an overpopulation of the $J=1$ level due to the
large ratio of radiative rates ($A_{21}/A_{10}\simeq 10$).

Once the level populations have been computed, the line spectra are
produced. The spectra are centered on the origin of the object. First the
circular images are produces in a range of frequencies. This circular
rendering of the images ensures that no details at large or small radii are
missed, and thus that no flux is accidently lost. The antenna temperatures
are then computed by integrating the images, after they have been multiplied
by the beam pattern centered on the origin of the object. We use an ``Airy''
beam, with a beam size corresponding to a single dish of 15 m diameter. The
object is placed at 140 parsec distance. The spectra of the four models,
computed for the first four radiative transitions at three different
inclinations, are shown in Fig.~\ref{fig-hcb-linespec}.

From the spectra one can clearly see the effect of flattening of the HCB
cloud, in particular for models 3 and 4.  At near pole-on inclination
(5$^{\circ}$) hardly any self-absorption is seen in these models, because
one looks straight into the ``cavity''. At near edge-on inclination
(85$^{\circ}$) the ``torus'' blocks the central regions from view near line
center, which results in the clear self-absorption features seen in the line
shapes. A similar manifestation of the non-spherical symmetry of the cloud
has been discussed recently by Van der Tak et
al.~\cite{vdtakdishoeck:1999}. An interesting feature of the line spectra
of models 3 and 4 is that the edge-on line profiles are wider than the
pole-on profiles. This can be attributed to the fact that the density and
the excitation temperature is lower at the pole than at the equator. At
pole-on inclination the high density equatorial matter will emit near
line-center instead of in the line wings, thus making the line profile
narrower.

The asymmetry between the red-shifted and the blue-shifted peaks are
typical for protostellar collapse. The rotation is hardly seen in
these spectra. This is because the rotational velocity is everywhere
much smaller than the free-fall velocity, except at very small radii
where the emission barely contributes to the single-dish spectra
shown here.

\ifthenelse{\figs=0}{}{
\begin{figure}
\mbox{}
\centerline{
\resizebox{3.6cm}{3.6cm}{\includegraphics{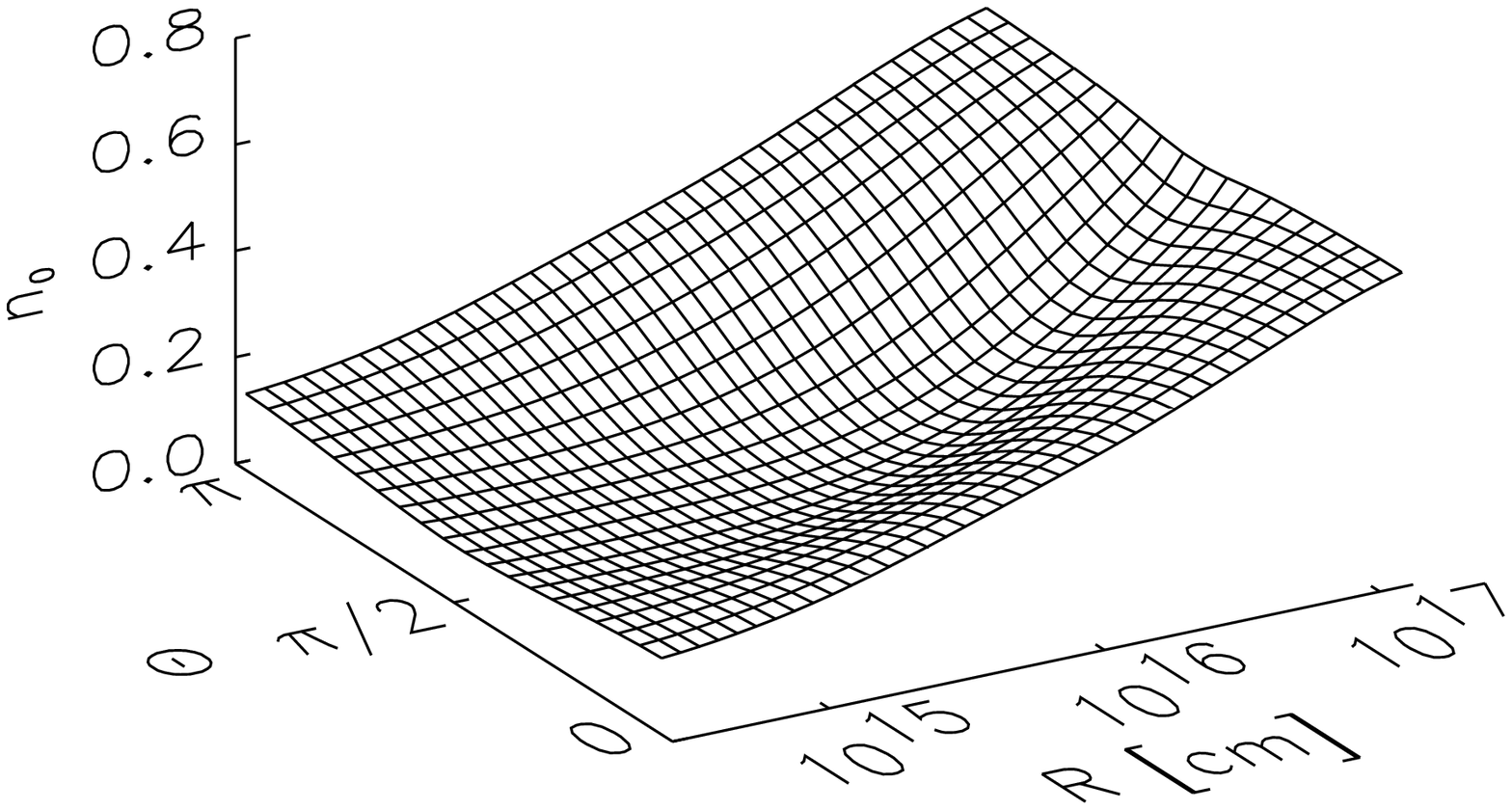}}\hspace{1em}
\resizebox{3.6cm}{3.6cm}{\includegraphics{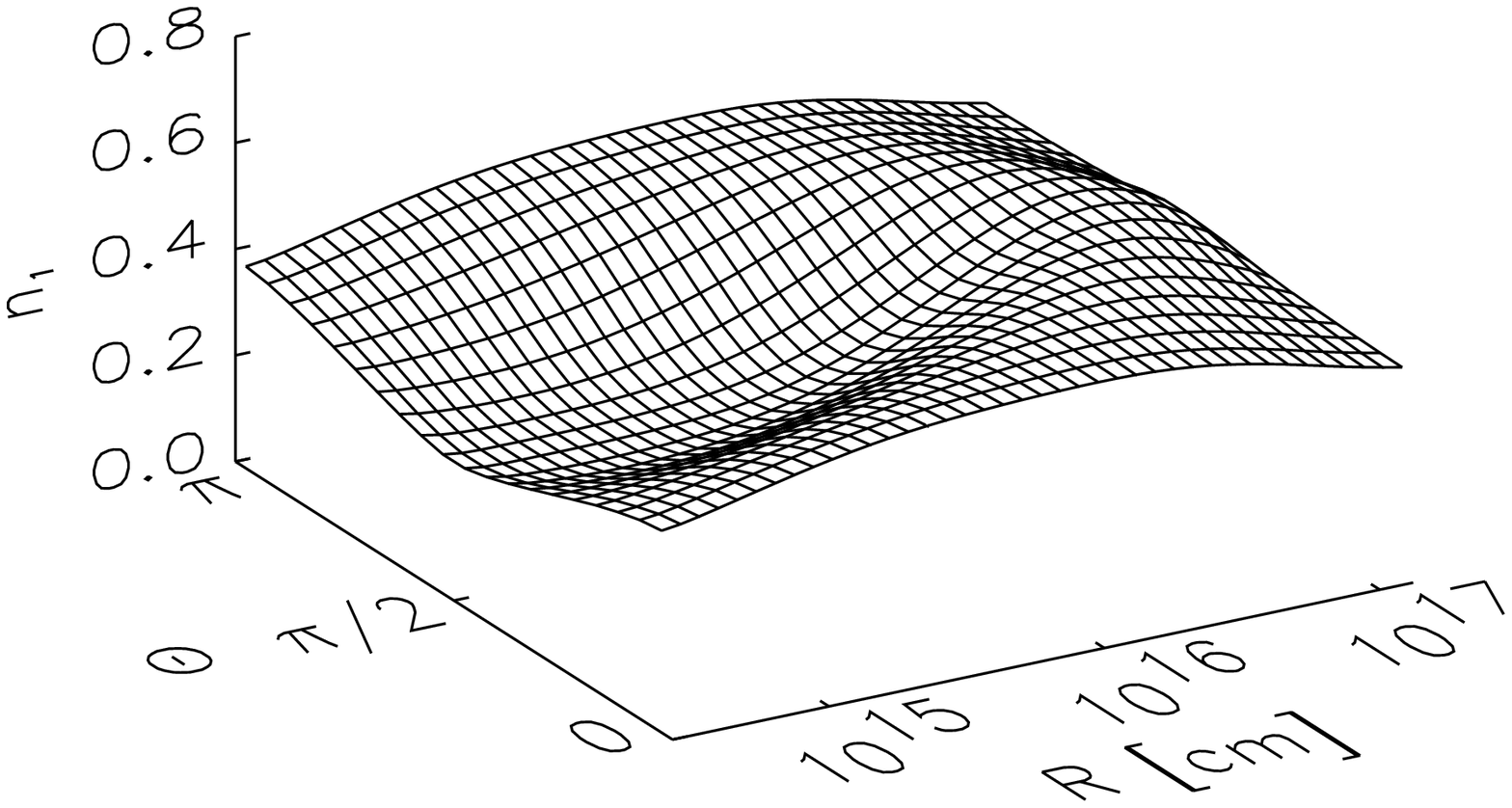}}
}
\centerline{
\resizebox{3.6cm}{3.6cm}{\includegraphics{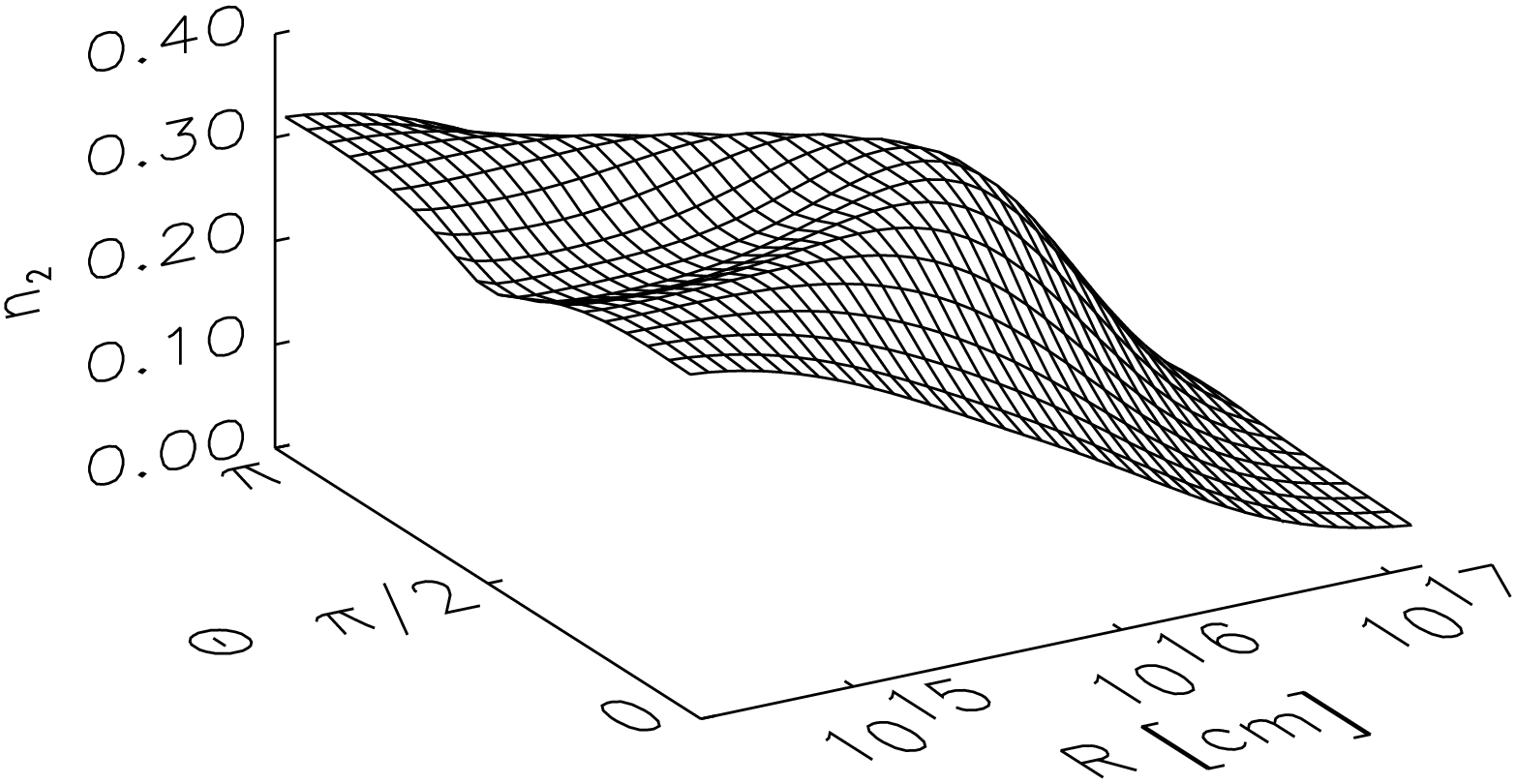}}\hspace{1em}
\resizebox{3.6cm}{3.6cm}{\includegraphics{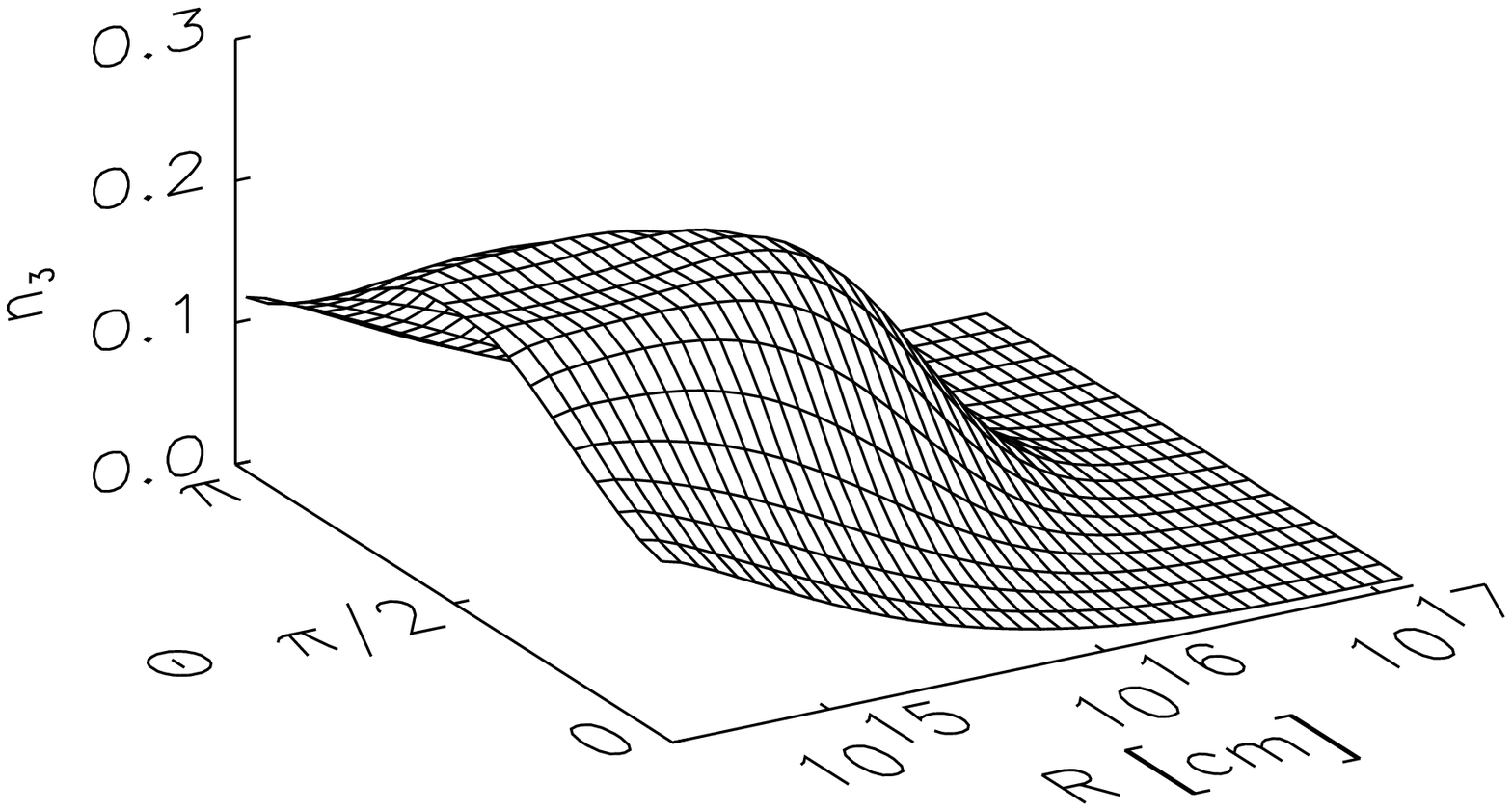}}
}
\mbox{}
\caption{The level populations for model 3 for the first four rotational
levels of HCO$^{+}$ as a function of $\Theta$ and $R$. Note that the
centrifugal radius (which is the disk outer edge) is at $\Rcentr=1.5\times
10^{15}\cm$.}
\label{fig-hcb-levpop}
\end{figure}
}
\ifthenelse{\figs=0}{}{
\begin{figure}
\mbox{}
\centerline{
\resizebox{3.6cm}{3.6cm}{\includegraphics{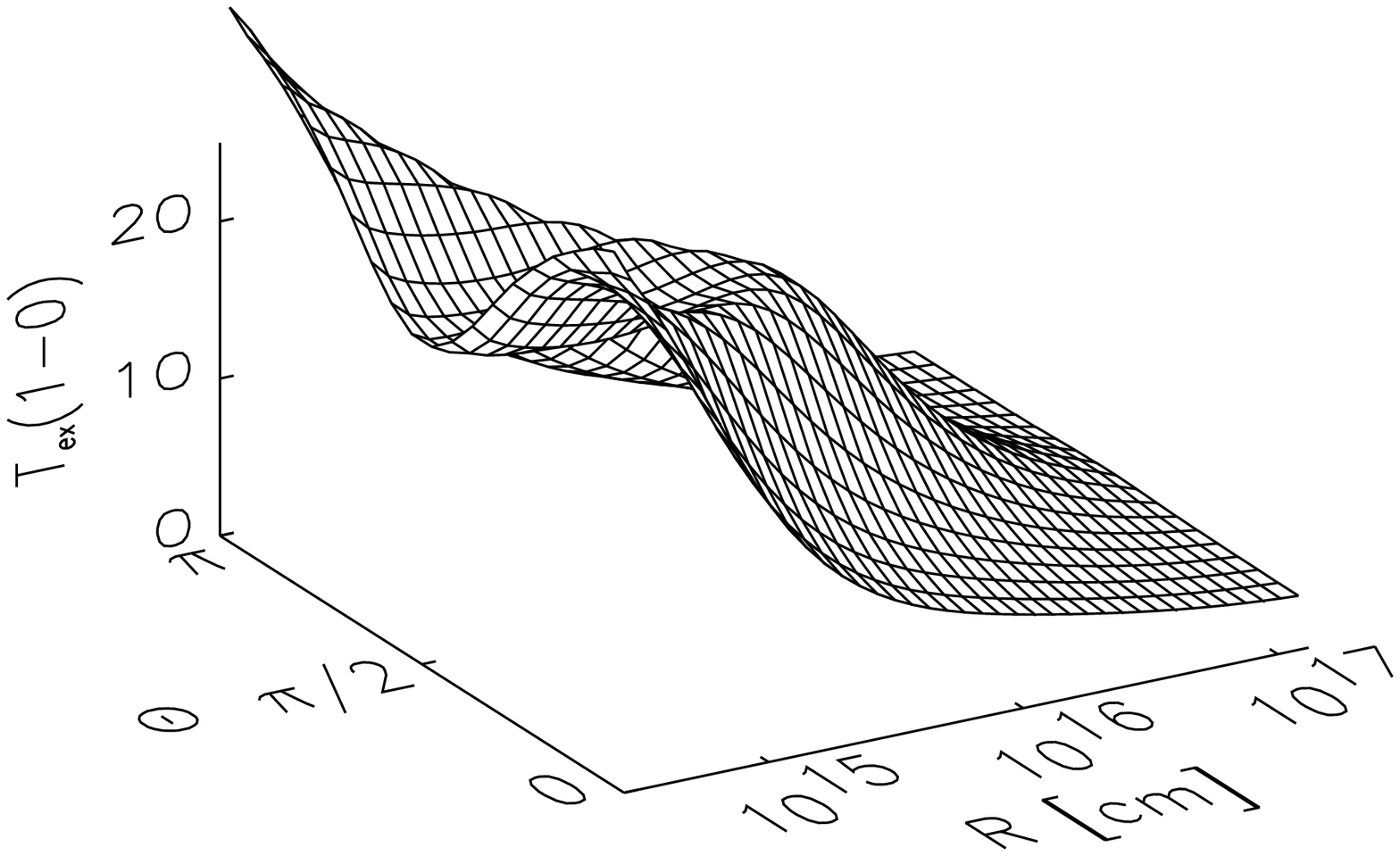}}\hspace{1em}
\resizebox{3.6cm}{3.6cm}{\includegraphics{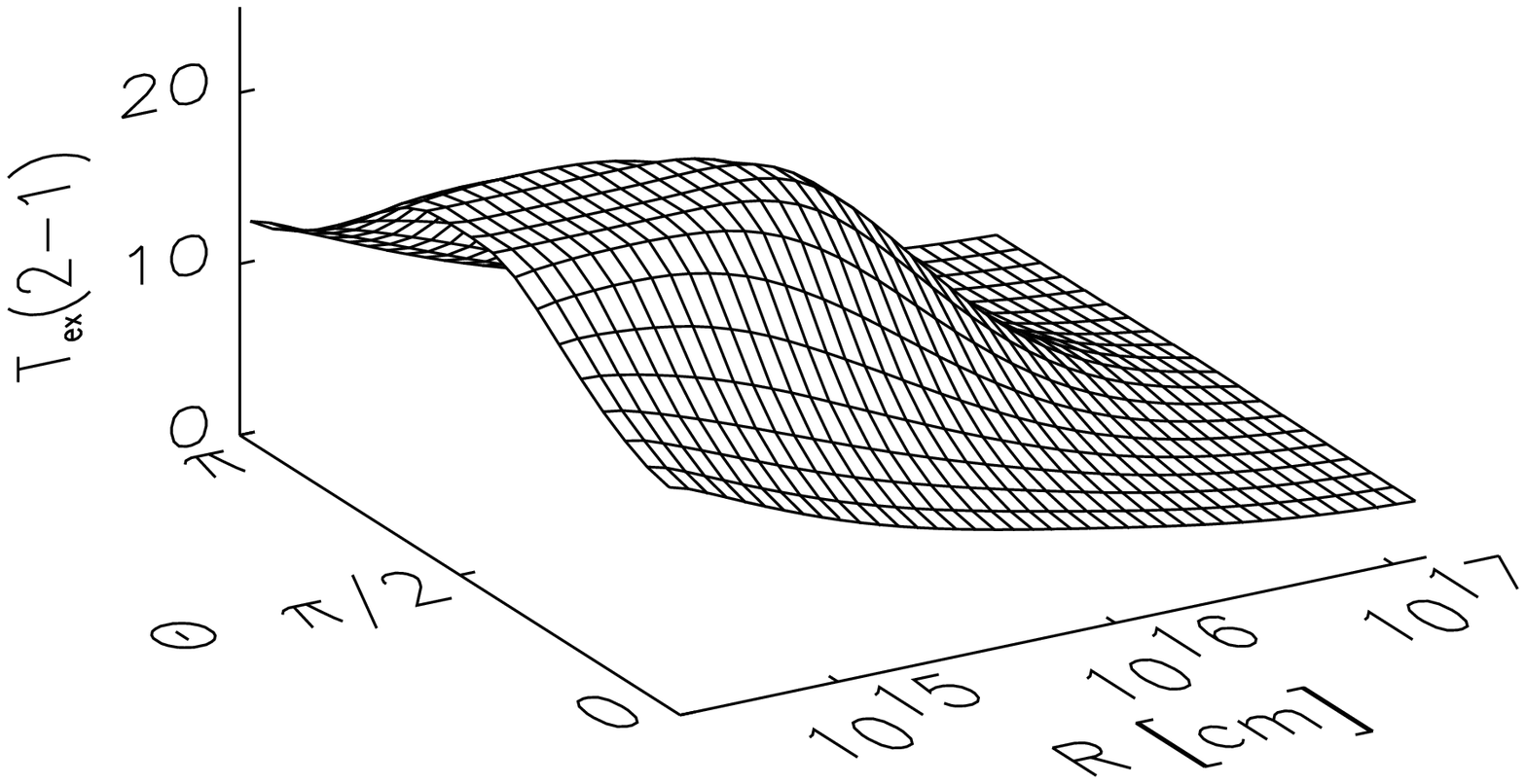}}
}
\centerline{
\resizebox{3.6cm}{3.6cm}{\includegraphics{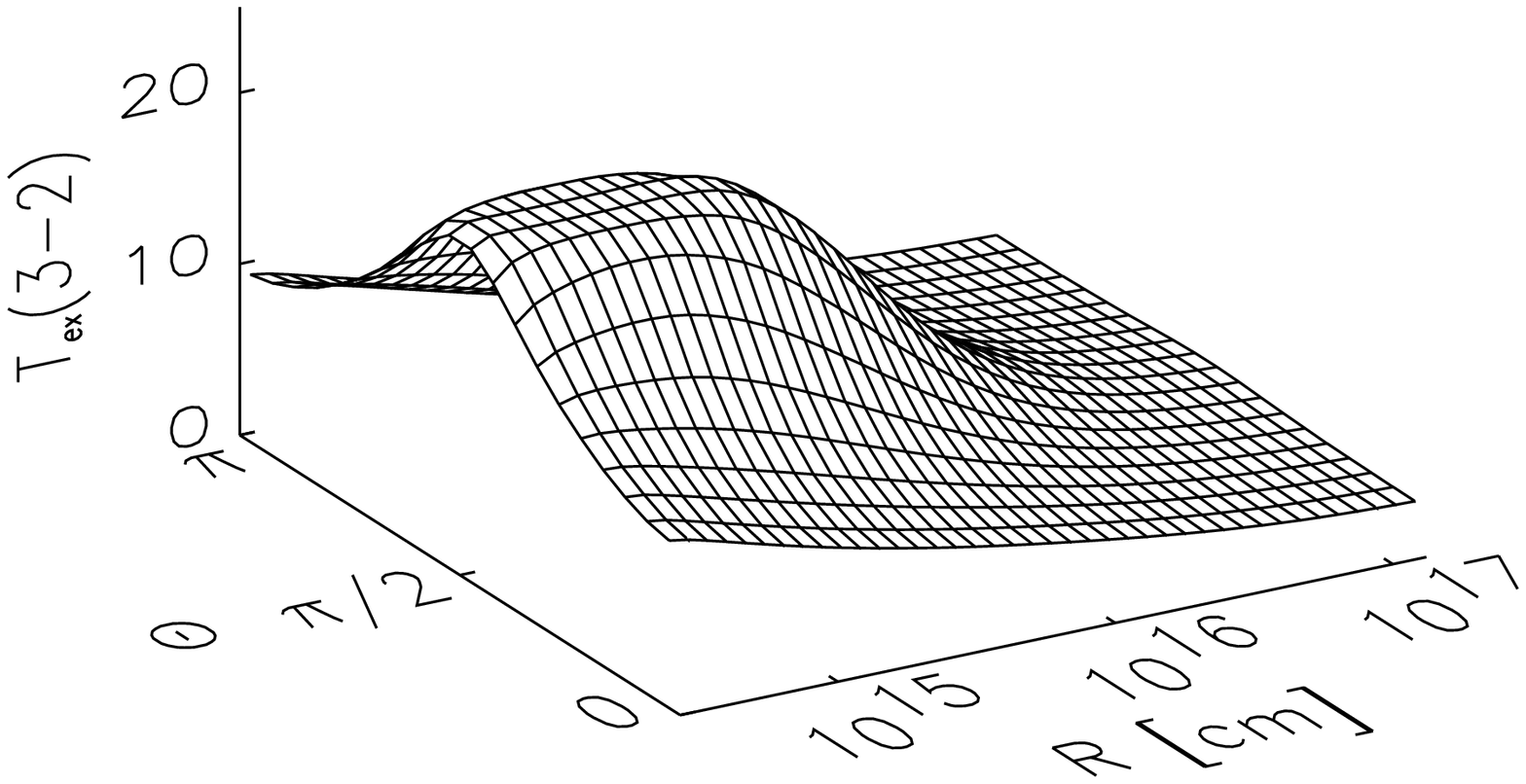}}\hspace{1em}
\resizebox{3.6cm}{3.6cm}{\includegraphics{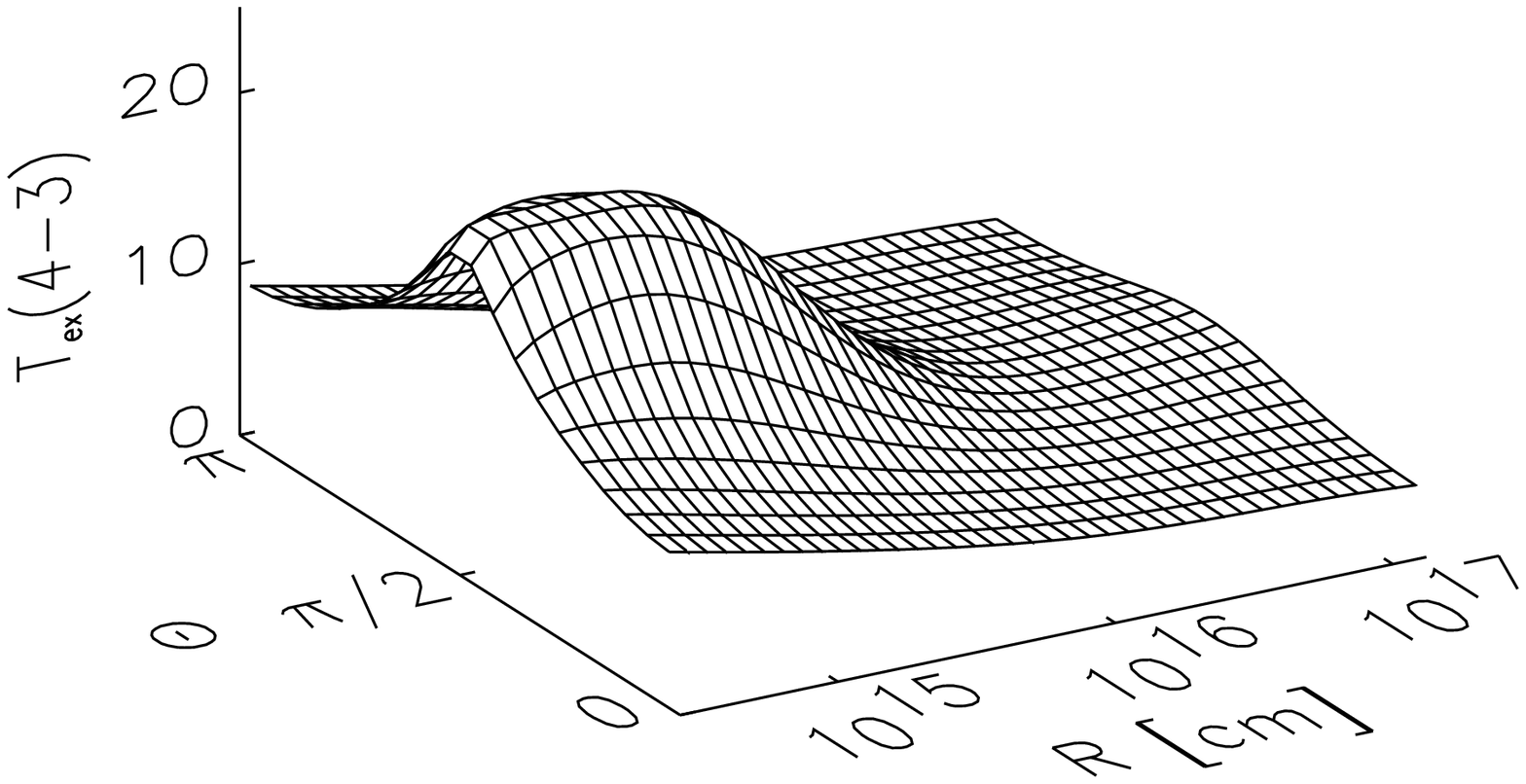}}
}
\mbox{}
\caption{The excitation temperatures of the lowest four rotational
transitions of HCO$^{+}$ for model 3, as a function of
$\Theta$ and $R$. They were deduced from the level populations shown
in Fig.~\ref{fig-hcb-levpop}, using the formula
$T_{\hbox{\scriptsize ex}(ij)}=(h\nu_{ij}/k)/\log(n_jg_i/n_ig_j)$.}
\label{fig-hcb-exctemp}
\end{figure}
}
\ifthenelse{\figs=0}{}{
\begin{figure*}
\mbox{}
\centerline{
\resizebox{14cm}{14cm}{\includegraphics{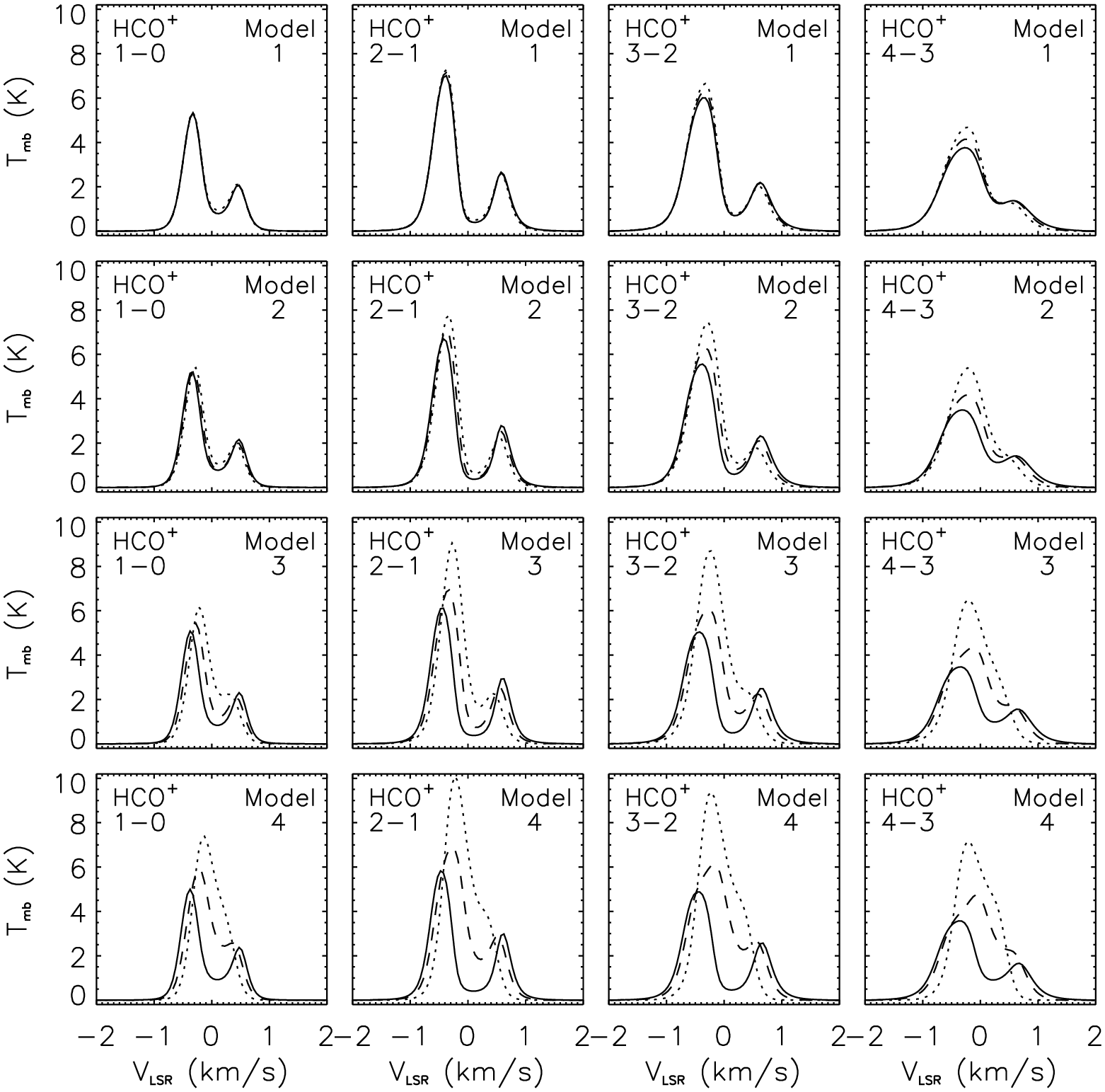}}
}
\vspace{2em}
\mbox{}
\caption{The single-dish HCO$^{+}$ line spectra of the HCB models, shown at
three different inclinations: dotted line is 5$^\circ$ (near pole-on),
dashed line is 45$^\circ$ and solid line is 85$^\circ$ (near edge-on). A
dish diameter of 15 m was taken, and the object is located at 140 pc.}
\label{fig-hcb-linespec}
\end{figure*}
}
\section{Conclusions}

Numerical radiative transfer modeling on desktop workstations is extremely
cheap. {\em Not} doing so in cases where this is possible would mean an
enormous waste of valuable information that lies encoded in observed
data. However the success of such modeling depends on the algorithms that
are available. We have developed a robust and accurate method, called the
``extended short characteristics'' (ESC) method, by which complicated 2--D
axi--symmetric multi--frequency radiative transfer calculations can be
performed. By using spherical coordinates, this method can accurately treat
circumstellar envelopes and disks from the stellar surface all the way up to
parsec scale, without the need of grid refinement. By making a special
choice of discrete photon angles and bundling `almost--radially--moving'
rays into a single bin, the conservation of radial flux can be guaranteed
even over many orders of magnitude in radius, and without excessive
computational cost. 

The ESC method, and a slight variation called MESC, forms the core of
a multi--purpose 2-D radiative transfer code called RADICAL.
We have tested the ESC/MESC algorithm on a simple test problem which we
described in this paper. We have also verified that the 2-D algorithm, when
applied to a 1-D spherically symmetric problem, indeed reproduces what an
independent 1-D algorithm would produce for the same problem. The errors
remained within a few percent, the exact value of which depends on the grid
resolution.

The ESC/MESC algorithm is designed for a variety of applications.  We have
demonstrated in this paper how the method can be used for the problem of
dust scattering in a bipolar proto--planetary nebula and the problem of
non-LTE line transfer in a collapsing cloud. But the method can also easily
be applied to other radiative processes, such as dust continuum emission
with radiative equilibrium for the dust grains, thermal Bremsstrahlung,
electron scattering and even Comptonization in hot plasmas. The ESC/MESC
algorithms are therefore a useful tool for a wide range of astrophysical
problems.

\section*{Acknowledgments}
We wish to thank Vincent Icke for making us aware of the applicability
of our algorithm (which was originally intended only for
Comptonization) to non--LTE line processes and dust absoption/emission
in stellar nebulae and winds, and for suggesting the Egg Nebula
application. We are grateful to Volker Ossenkopf for sending us his
solution to the test problem.
CPD thanks Gerd--Jan van Zadelhof for useful discussions
on line transfer, Ewine van Dishoeck, Rens Waters and Alex de Koter
for discussions on the applicability of the algorithm to envelopes
around young and old stars. CPD thanks Bj\"orn Heijligers and Vincent
de Heij for their help with IDL. Special thanks to Jeremy Yates,
Floris v.d.~Tak and Volker Ossenkopf for their collaboration
in comparing the results of RADICAL with those of their codes 
for molecular line transfer.


\begin{thebibliography}{}

\bibitem[\protect\astroncite{Auer et~al.}{1994}]{auerfabtruj:1994}
Auer, L., Bendicho, P.~F., and Bueno, J.~T., 1994,
\newblock {\aap} {292}, 599

\bibitem[\protect\astroncite{{Auer} and {Mihalas}}{1969}]{auermihalas:1969}
{Auer}, L.~H. and {Mihalas}, D., 1969,
\newblock {\apj} {158}, 641+

\bibitem[\protect\astroncite{Auer and Paletou}{1994}]{aupal:1994}
Auer, L.~H. and Paletou, F., 1994,
\newblock {\aap} {284}, 675

\bibitem[\protect\astroncite{{Bachiller}}{1996}]{bachiller:1996}
{Bachiller}, R., 1996,
\newblock {\araa} {34}, 111

\bibitem[\protect\astroncite{{Bieging} and {Q.-Rieu}}{1996}]{bieging:1996}
{Bieging}, J.~H. and {Q.-Rieu}, N., 1996,
\newblock {\aj} {112}, 706

\bibitem[\protect\astroncite{Busche and Hillier}{2000}]{buschehillier:2000}
Busche, J. and Hillier, D., 2000,
\newblock {\apj} {531}, 1071

\bibitem[\protect\astroncite{{Cassen} and {Moosman}}{1981}]{cassenmoosman:1981}
{Cassen}, P. and {Moosman}, A., 1981,
\newblock {Icarus} {48}, 353

\bibitem[\protect\astroncite{Collison and Fix}{1991}]{collfix:1991}
Collison, A. and Fix, J., 1991,
\newblock {\apj} {368}, 545

\bibitem[\protect\astroncite{{Crampton} et~al.}{1975}]{crampton:1975}
{Crampton}, D., {Cowley}, A.~P., and {Humphreys}, R.~M., 1975,
\newblock {\apjl} {198}, L135

\bibitem[\protect\astroncite{Dullemond}{1999}]{dullemondthesis:1999}
Dullemond, C.~P., 1999,
\newblock {Ph.D. thesis}, Univesiteit Leiden

\bibitem[\protect\astroncite{Efstathiou and
  Rowan-Robinson}{1991}]{efstrow:1991}
Efstathiou, A. and Rowan-Robinson, M., 1991,
\newblock {\mnras} {252}, 528

\bibitem[\protect\astroncite{{Galli} and {Shu}}{1993}]{gallishu:1993a}
{Galli}, D. and {Shu}, F.~H., 1993,
\newblock {\apj} {417}, 220

\bibitem[\protect\astroncite{{Green}}{1975}]{green:1975}
{Green}, S., 1975,
\newblock {\apj} {201}, 366

\bibitem[\protect\astroncite{{Hartmann} et~al.}{1996}]{hartcalvboss:1996}
{Hartmann}, L., {Calvet}, N., and {Boss}, A., 1996,
\newblock {\apj} {464}, 387

\bibitem[\protect\astroncite{Hogerheijde}{1998}]{hogerheijdethesis:1998}
Hogerheijde, M., 1998,
\newblock {Ph.D. thesis}, Rijks Univesiteit Leiden

\bibitem[\protect\astroncite{Hubeny}{1989}]{hubeny:1989}
Hubeny, I., 1989,
\newblock {in Theory of Accretion Disks, ed. F. Meyer et al.},
\newblock Kluwer

\bibitem[\protect\astroncite{Kunasz and Auer}{1988}]{kunauer:1988}
Kunasz, P.~B. and Auer, L.~H., 1988,
\newblock {\jqsrt} {39}, 67

\bibitem[\protect\astroncite{Latter et~al.}{1993}]{latterhora:1993}
Latter, W., Hora, J., Kelly, D., Deutsch, L., and P.R.Maloney, 1993,
\newblock {\aj} {106}, 260

\bibitem[\protect\astroncite{{Lenzen}}{1987}]{lenzen:1987}
{Lenzen}, R., 1987,
\newblock {\aap} {173}, 124

\bibitem[\protect\astroncite{Leung and Liszt}{1976}]{leunglist:1976}
Leung, C. and Liszt, H., 1976,
\newblock {\apj} {208}, 732

\bibitem[\protect\astroncite{{McCaughrean} and
  {O'Dell}}{1996}]{caughodell:1996}
{McCaughrean}, M.~J. and {O'Dell}, C.~R., 1996,
\newblock {\aj} {111}, 1977+

\bibitem[\protect\astroncite{Mihalas et~al.}{1975}]{mihkunhum:1975}
Mihalas, D., Kunasz, P.~B., and Hummer, D.~G., 1975,
\newblock {\apj} {202}, 465

\bibitem[\protect\astroncite{Mihalas et~al.}{1978}]{mam:1978}
Mihalas, D.~M., Auer, L.~H., and Mihalas, B.~R., 1978,
\newblock {\apj} {220}, 1001

\bibitem[\protect\astroncite{{Monteiro}}{1985}]{monteiro:1985}
{Monteiro}, T.~S., 1985,
\newblock {\mnras} {214}, 419

\bibitem[\protect\astroncite{Morris}{1981}]{morris:1981}
Morris, M., 1981,
\newblock {\apj} {249}, 572

\bibitem[\protect\astroncite{Murray et~al.}{1994}]{murraycas:1994}
Murray, S., Castor, J., Klein, R., and McKee, C., 1994,
\newblock {\apj} {435}, 631

\bibitem[\protect\astroncite{{Myers} et~al.}{1991}]{meyersfuller:1991}
{Myers}, P.~C., {Fuller}, G.~A., {Goodman}, A.~A., and {Benson}, P.~J., 1991,
\newblock {\apj} {376}, 561

\bibitem[\protect\astroncite{{Ney} et~al.}{1975}]{ney:1975}
{Ney}, E.~P., {Merrill}, K.~M., {Becklin}, E.~E., {Neugebauer}, G., and
  {Wynn-Williams}, C.~G., 1975,
\newblock {\apjl} {198}, L129

\bibitem[\protect\astroncite{Olson and Kunasz}{1987}]{olsenkunasz:1987}
Olson, G. and Kunasz, P., 1987,
\newblock {\jqsrt} {38}, 325

\bibitem[\protect\astroncite{Ossenkopf}{1999}]{ossenkopfsimline:1999}
Ossenkopf, V., 1999,
\newblock {http://waww.ph1.uni-koeln.de/\~{}ossk/Myself/simline.html}

\bibitem[\protect\astroncite{Philips}{1999}]{rphilips:1999}
Philips, R., 1999,
\newblock {Ph.D. thesis}, University of Kent

\bibitem[\protect\astroncite{Rutten}{1999}]{rutten:1999}
Rutten, R., 1999,
\newblock {Radiative Transfer in Stellar Atmospheres},
\newblock http://www.fys.ruu.nl/\~{}rutten/

\bibitem[\protect\astroncite{Rybicki and Hummer}{1991}]{rybhum:1991}
Rybicki, G. and Hummer, D., 1991,
\newblock {\aap} {245}, 171

\bibitem[\protect\astroncite{Sahai et~al.}{1998}]{sahaitraug:1998}
Sahai, R., Trauger, J., Watson, A., and et~al., K.~S., 1998,
\newblock {\apj} {493}, 301

\bibitem[\protect\astroncite{{Scharmer}}{1981}]{scharmer:1981}
{Scharmer}, G.~B., 1981,
\newblock {\apj} {249}, 720

\bibitem[\protect\astroncite{{Shu}}{1977}]{shu:1977}
{Shu}, F.~H., 1977,
\newblock {\apj} {214}, 488

\bibitem[\protect\astroncite{Sonnhalter et~al.}{1995}]{sonnpreiyo:1995}
Sonnhalter, C., Preibisch, T., and Yorke, H., 1995,
\newblock {\aap} {299}, 545

\bibitem[\protect\astroncite{Spaans}{1996}]{spaans:1996}
Spaans, M., 1996,
\newblock {\aap} {307}, 271

\bibitem[\protect\astroncite{Stone et~al.}{1992}]{stonemihnor:1992}
Stone, J., Mihalas, D., and Norman, M., 1992,
\newblock {\apjs} {80}, 819

\bibitem[\protect\astroncite{{Terebey} et~al.}{1984}]{terebyshucas:1984}
{Terebey}, S., {Shu}, F.~H., and {Cassen}, P., 1984,
\newblock {\apj} {286}, 529

\bibitem[\protect\astroncite{{Ulrich}}{1976}]{ulrich:1976}
{Ulrich}, R.~K., 1976,
\newblock {\apj} {210}, 377

\bibitem[\protect\astroncite{van~der Tak et~al.}{1999}]{vdtakdishoeck:1999}
van~der Tak, F., van Dishoeck N.J.~Evans, E., Bakker, E., and Blake, G., 1999,
\newblock {ApJ} {522},

\bibitem[\protect\astroncite{{Wolf} et~al.}{1999}]{wolfhenning:1999}
{Wolf}, S., {Henning}, T., and {Stecklum}, B., 1999,
\newblock {\aap} {349}, 839

\bibitem[\protect\astroncite{{Yorke} et~al.}{1993}]{yorkebodlau:1993}
{Yorke}, H.~W., {Bodenheimer}, P., and {Laughlin}, G., 1993,
\newblock {\apj} {411}, 274

\bibitem[\protect\astroncite{{Zane} et~al.}{1996}]{zanetur:1996}
{Zane}, S., {Turolla}, R., {Nobili}, L., and {Erna}, M., 1996,
\newblock {\apj} {466}, 871

\bibitem[\protect\astroncite{{Zhou}}{1992}]{zhou:1992}
{Zhou}, S., 1992,
\newblock {\apj} {394}, 204

\end{thebibliography}
\end{document}